\documentclass[a4paper,11pt]{article}

\usepackage{amsmath, amssymb, bm}
\usepackage[top=1in, bottom=1.2in, left=0.7in, right=0.7in]{geometry}
\usepackage[]{graphicx}
\usepackage[font=scriptsize,labelfont=bf]{caption}
\usepackage[caption=false]{subfig}
\usepackage{slashed}
\usepackage{textcomp}
\usepackage{authblk}
\usepackage{hyperref}
\newcommand{\be}{\begin{eqnarray}}
\newcommand{\ee}{\end{eqnarray}}

\newcommand{\bs}{\boldsymbol}

\begin{document}

\title{\bf{GRAPH mixing}}
\author{Damian Ejlli and Venugopal R. Thandlam}

\affil{\emph{\normalsize{Department of Physics, Novosibirsk State University, Novosibirsk 630090, Russia}}}

\date{}

\maketitle

\begin{abstract}

In the era of gravitational wave (GW) detection from astrophysical sources by LIGO/VIRGO, it is of great importance to take the quantum gravity effect of graviton-photon (GRAPH) mixing in the cosmic magnetic field to the next level. In this work, we study such an effect and derive for the first time perturbative solutions of the linearized equations of motions of the GRAPH mixing in an expanding universe. In our formalism we take into account all known standard dispersive and coherence breaking effects of photons such as the Faraday effect, the Cotton-Mouton (CM) effect, and the plasma effects in the cosmic magnetic field. Our formalism, applies to a cosmic magnetic field either a uniform or a slowly varying non-homogeneous field of spacetime coordinates with an arbitrary field direction. For binary systems of astrophysical sources of GWs at extragalactic distances with chirp masses $M_\text{CH}$ of a few solar masses, GW present-day frequencies $\nu_0\simeq 50-700$ Hz, and present-day cosmic magnetic field amplitudes $\bar B_0\simeq 10^{-10}-10^{-6}$ G, the power of electromagnetic radiation generated in the GRAPH mixing at present is substantial and in the range $P_\gamma\simeq 10^6-10^{15}$ (erg/s). On the other hand, the associated power flux $F_\gamma$ is quite faint depending on the source distance with respect to the Earth. Since in the GRAPH mixing the velocities of photons and gravitons are preserved and are equal, this effect is the only one known to us, whose certainty of the contemporary arrival of GWs and electromagnetic radiation at the detector is guaranteed.

\end{abstract}

\vspace{1cm}

\section{Introduction}
\label{sec:1}

The detections of several GW events by the LIGO/VIRGO Collaborations \cite{Abbott:2016blz}, have finally confirmed a long-standing problem, that indeed spacetime perturbations that propagate with the speed of light and that are not an artifact prediction of the theory of general relativity do exist. The detection of GWs followed after several decades of intensive theoretical studies and experimental efforts that took a great push forward starting from the first detection of a GW source, namely the PSR B1913+16 binary system of neutron stars \cite{Hulse:1974eb}.  The LIGO/VIRGO detections apart from being important in many aspects of physics shed a new light in favor of the graviton, namely the quantized particle of spin two of the gravitational field. The GW events detected by the LIGO/VIRGO Collaborations, so far, have confirmed with good accuracy that GWs propagate in the vacuum with the speed of light and if the graviton is a massive particle, its mass should be smaller than $m_g<1.2\times 10^{-22}$ eV; see Refs. \cite{Abbott:2016blz} for details.

One of the key assumptions about the nature of GWs is that they weakly interact with matter and fields while propagating from the source to the detector, and consequently their velocities and amplitudes are assumed to remain unaltered. This assumption is justifiable in most situations because being the interaction strength of GWs with matter and fields very small, one usually does not expect any loss or transformation of GWs propagating though cosmological distances. Even though this assumption is quite realistic in most cases, there might be some exceptions in the case when GWs interact with spatially extended electromagnetic fields comparable with astrophysical and cosmological distances. Indeed, as the theory of general relativity teaches us, every form of nonstationary stress energy-tensor on the right-hand side of the Einstein field equations with a quadrupole moment produces spacetime perturbations or simply GWs.  So, in principle nonstationary interactions among electromagnetic fields would produce GWs.

While nonstationary interactions among electromagnetic fields with quadrupole moments produce GWs such as the interaction of a plane electromagnetic wave with a static magnetic field, it is also possible that the interaction of GWs with external electromagnetic fields would produce electromagnetic radiation out of GWs. Therefore, the overall outcome is that GWs and electromagnetic waves mix with each other in the presence of external electromagnetic fields, and this effect propagates in space throughout the region where the external electromagnetic field is spatially located; see Ref. \cite{Ejlli:2013gaa} for an intuitive explanation. 
Based on this fundamental prediction of the theory of general relativity, the possibility to generate GWs in the laboratory from the interaction of electromagnetic radiation with external prescribed static magnetic fields was initially proposed in Ref.  \cite{gertsen61}.

 Through the decades the possibility of mixing GWs with electromagnetic waves and vice versa in a constant external magnetic field has been studied by several authors \cite{Boccaletti70}-\cite{Bastianelli05} for some specific magnetic field configuration, which in most cases has been taken to be perpendicular to the propagation of the incident GW and/or electromagnetic wave. In those cases where the field was not taken to be perpendicular with respect to the incident field propagation, important dispersive and coherence breaking effects such as the Faraday effect and the CM effect have not been taken into account. In these studies, classical, semi-classical \cite{Boccaletti70} and field theory approaches \cite{Bastianelli05} have been employed to the mixing problem, and some possibilities for applying this effect in cosmological scenarios have been proposed in Ref. \cite{Dolgov:2012be}. A different way to produce electromagnetic waves due to propagation of GWs in vacuum has been proposed in Ref. \cite{Jones:2017dzt}.

 In order for the GRAPH mixing to work, it is necessary to have an external electromagnetic field and in cosmological situations it can be possible in the presence of large-scale cosmic magnetic fields (for general concepts on cosmic magnetic fields see Ref. \cite{Grasso:2000wj}).  Indeed, as it is well known, the presence of large-scale magnetic field in galaxies and galaxy clusters has been experimentally verified, while it is still unclear if such field is present in the intergalactic space. In galaxy clusters, the measurements of the rotation angle of the received light due to the Faraday effect confirm the presence of a large-scale magnetic field inside them, with a magnitude of the order of a few $\mu$G. On the other hand, in the intergalactic space recent studies by the Planck collaboration \cite{Ade:2015cva} would suggest a weaker large scale cosmic magnetic field with upper limit field strength $\bar B_0 \lesssim 3 - 1380$ nG at the correlation length scale $\lambda_B=1$ Mpc. The limit of the order of 1380 nG is set from the Faraday effect of the CMB, while the limit of $\bar B_0\lesssim 3$ nG is set from the CMB temperature anisotropy. In addition, from the non observation of gamma ray emission from the intergalactic medium due to the injection of high energy particles by blazars \cite{Neronov:1900zz}, a lower value on the strength of the intergalactic magnetic field of the order  $\bar B_0\geq 10^{-16}-10^{-15}$ G is inferred.

 The detection of GWs from astrophysical binary systems gives a rather unique opportunity to probe the GRAPH mixing effect in the cosmic magnetic field. Some important questions that we can ask at this stage are the following; If large-scale magnetic fields do exist, what is the probability of transformation of GWs into electromagnetic radiation? What is the energy per unit time and/or the energy density received at the Earth? What is the polarization of the electromagnetic radiation received? In this work, we address these questions by applying the GRAPH mixing to astrophysical binary systems located at extragalactic distances (not located in our galaxy) with redshifts $0.1\lesssim z$, and we make predictions for the energy power and energy power flux of the electromagnetic radiation generated in the GRAPH mixing. With respect to other works where the GRAPH mixing was studied for constant magnetic field  \cite{Boccaletti70}-\cite{Bastianelli05} in a laboratory and in the early universe where the density matrix equations of motions were solved numerically \cite{Dolgov:2012be}, in this work we find analytic solutions of the field equations of motion for a slowly varying non-homogeneous magnetic. In addition, with respect to other studies \cite{Boccaletti70}, \cite{Bastianelli05}, \cite{Dolgov:2012be} we allow the direction of the external magnetic field to be arbitrary with respect to the GW direction of propagation and take into account the Faraday and CM effects in the magnetic field.
 
 This paper is organized as follows: In Sec. \ref{sec:2} we derive the linearized field equations of motion in a spatially and temporally non-homogeneous magnetic field with the field inhomogeneity scale bigger than the GW wavelength. In Sec. \ref{sec:3} we discuss all standard dispersive and coherence breaking electromagnetic wave effects in a magnetized plasma by writing explicitly the elements of the photon polarization tensor in a magnetized medium. In Sec. \ref{sec:4} we find analytic solutions of the linearized equations of motion by using perturbation theory. In Sec. \ref{sec:5} we find the Stokes parameters of the electromagnetic radiation generated in the GRAPH mixing. In Sec. \ref{sec:6} we find some analytic expressions of the integrals that do appear in the Stokes parameters. In Sec. \ref{sec:7} we calculate the power and the power flux of the electromagnetic radiation generated in the GRAPH mixing. In Sec. \ref{sec:8} we discuss possible cutoffs in the GRAPH spectrum due to plasma frequency, and in Sec. \ref{sec:9} we conclude. In this work we use the metric with signature $\eta_{\mu\nu}=\text{diag}[1, -1, -1, -1]$ and work with the rationalized Lorentz-Heaviside natural units ($k_B=\hbar=c=\varepsilon_0=\mu_0=1$) with $e^2=4\pi \alpha$. In addition, in this work we use the values of the cosmological parameters found by the Planck Collaboration \cite{Ade:2015xua}with $\Omega_\Lambda \simeq 0.68, \Omega_\text{M}\simeq 0.31, h_0\simeq 0.67$ with zero spatial curvature $\Omega_\kappa=0$.

 \section{Field mixing in external magnetic field}
 \label{sec:2}
 
To describe the GRAPH mixing, it is necessary first to start with the total action of the GRAPH mixing. In general, the action for a given Lagrangian density $\mathcal L$ minimally coupled to gravity is $S=\int d^4x \sqrt{-g}\,\mathcal L$ where $\mathcal L$  describes the total Lagrangian density of matter and fields and their interactions. In our case, it is given by the sum of the following terms
 \begin{equation}
\mathcal L=\mathcal L_\text{gr}+\mathcal L_\text{em},
\end{equation}
where $\mathcal L_\text{gr}$ and $\mathcal L_\text{em}$ are, respectively, the Lagrangian densities of  gravitational and electromagnetic fields. These terms are, respectively, given by
\begin{align}\label{lag-dens}
\mathcal L_\text{gr} &=\frac{1}{\kappa^2}\,R,\quad \mathcal L_\text{em}=-\frac{1}{4}F_{\mu\nu}F^{\mu\nu}-\frac{1}{2}\int d^4x^\prime A_\mu(x) \Pi^{\mu\nu}(x, x^\prime)A_\nu(x^\prime).
\end{align}
Here $R$ is the Ricci scalar, $g$ is the metric determinant, $F_{\mu\nu}$ is the electromagnetic field tensor, $\kappa^2=16\pi G_\text{N}$ with $G_\text{N}$ being the Newtonian constant and $\Pi^{\mu\nu}$ is the photon  polarization tensor in a magnetized medium.

 By expanding the metric tensor around the flat Minkowski spacetime as $g_{\mu\nu}=\eta_{\mu\nu}+\kappa h_{\mu\nu}+...$, we get the following expression for the total effective action: 
\begin{align}\label{tot-lang}
\mathcal S_\text{eff} & = \frac{1}{4}\int d^4 x \left[2\partial_\mu h^{\mu\nu}\partial_\rho h_\nu^\rho+\partial_\mu h\partial^\mu h-\partial_\mu h_{\alpha\beta}\partial^\mu h^{\alpha\beta}-2\partial_\mu h^{\mu\nu}\partial_\nu h\right]-\frac{1}{4}\int d^4 x F_{\mu\nu}F^{\mu\nu}+ \frac{\kappa}{2}\int d^4x h_{\mu\nu} T_\text{em}^{\mu\nu}\nonumber\\ & -\frac{1}{2}\int d^4 x\int d^4x^\prime A_\mu(x) \Pi^{\mu\nu}(x, x^\prime)A_\nu(x^\prime) + O(\kappa \partial h^3)+O(\kappa h \Pi),
\end{align}
where $h_{\mu\nu}$ is the gravitational wave tensor with $h=\eta_{\mu\nu}h^{\mu\nu}$ and $T_\text{em}^{\mu\nu}$ is the electromagnetic field tensor \footnote{With the metric with signature $\eta_{\mu\nu}=\text{diag}[1, -1, -1, -1]$, the expressions for the spatial components of the electromagnetic stress-energy tensor are $T_{ij}=\mathcal E_i \mathcal E_j + \mathcal B_i \mathcal B_j - (1/2)\delta_{ij}(\mathcal E^2+\mathcal B^2)$ where $\mathcal E_i=E_i+\bar E_i, \mathcal B_i=B_i+\bar B_i$ are respectively the components of the total electric and magnetic fields. The stress-energy tensor of the incident photon field tensor, $f_{\mu\nu}$, is not a source of GWs; see Ref. \cite{Boccaletti70} for details.}.

\begin{figure}[htbp]
\begin{center}
\includegraphics[scale=0.8]{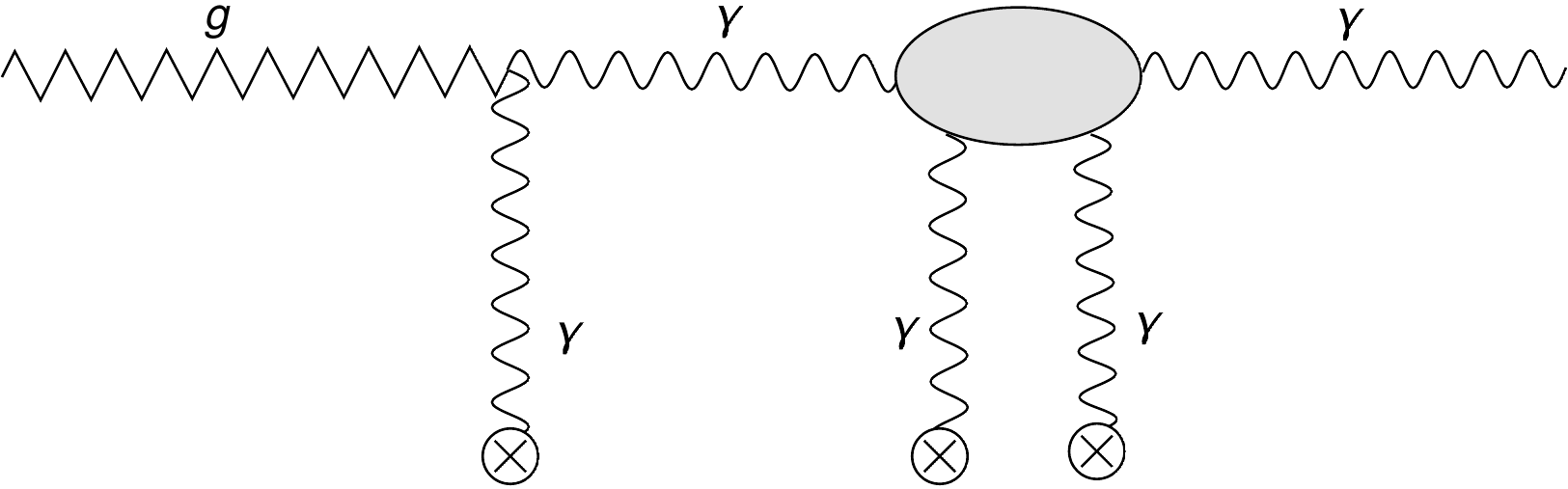}
\caption{Typical Feynman diagram for the GRAPH mixing in external magnetic field. The zigzag line denotes a graviton, the wavy lines denote photons, and the cross vertexes denote the external magnetic field. Here we have also included the photon self-energy or photon polarization tensor $\Pi_{\mu\nu}$ in a magnetized medium that is represented by the grey loop.}
\label{Graphax}
\end{center}
\end{figure}

Let us suppose that we have GWs propagating in a vacuum and after they enter a region where only an external magnetic field exists. We can put GWs in the TT gauge \emph{before} entering the magnetic field region, namely $h_{0i}=0, \partial^j h_{ij}=0,h_i^i=0$. The Euler-Lagrange equations of motion from \eqref{tot-lang} for the propagating photon and graviton fields components, $A^\mu$ and $h_{ij}$ propagating in the external magnetic field, are given by 
\begin{eqnarray}
\nabla^2 A^0 &=& 0,\nonumber\\ 
\Box \bs A^i +\left(\int d^4x^\prime \Pi^{i j}(x, x^\prime)\bs A_j(x^\prime)\right)+ \partial^i\partial_\mu A^\mu &=& \kappa\,\partial_\mu[h^{\mu\beta}\bar {F}_{\beta}^{i}-h^{i\beta}\bar {F}_{\beta}^{\mu}],\nonumber\\
\Box h_{ij} &=&-\kappa\, (B_i\bar B_j+\bar B_i B_j +\bar B_i\bar B_j).
 \label{system1}
\end{eqnarray}

In obtaining the system of Eqs. \eqref{system1}, the electromagnetic field tensor has been written as the sum of the incident photon field tensor $f_{\mu\nu}$ and of the external field tensor $\bar F_{\mu\nu}$, namely $F_{\mu\nu}=f_{\mu\nu}+\bar F_{\mu\nu}$. However, since we are assuming that only an external magnetic field exists, we essentially have only that $\bar F_{ij}\neq 0$. In addition, we assume that the external magnetic field varies in space on much larger scales than the incident GW wavelength, namely $\lambda_B\gg \lambda_\text{gw}$. The latter assumption does not necessarily mean that the external magnetic field is only a uniform function of space coordinates where the condition $\lambda_B\gg \lambda_\text{gw}$ is always satisfied. In contrast, the magnetic field is assumed to be a slowly varying function of space coordinates;  namely the field could be as well non homogeneous in space and in time as well. The condition $\lambda_B\gg \lambda_\text{gw}$ implies that $|h\partial \bar F|\ll |\bar F \partial h|$, where for simplicity we suppressed the indices in $h_{ij}$ and $\bar F_{ij}$. By using these approximations, we can simplify the system \eqref{system1} and write it in the form
\begin{eqnarray}
\nabla^2 A^0 &=& 0 ,\nonumber\\
 \Box \bs A^i +\left(\int d^4x^\prime \Pi^{i j}(x, x^\prime)\bs A_j(x^\prime)\right)+ \partial^i\partial_\mu A^\mu &=&  - \kappa\,(\partial_j h^{ik})\bar F_{k}^j,\nonumber\\
\Box h_{ij} &=&-\kappa\left(B_i\bar B_j+\bar B_i B_j +\bar B_i\bar B_j \right), 
  \label{system2}
\end{eqnarray}
where we used the fact that $F_{\mu\nu}\tilde F^{\mu\nu}=-4 \bs E\cdot \bs B$, $\tilde F^{0i}=-\bs B^i$ and we used the TT-gauge conditions.

To solve the system of Eqs. \eqref{system2}, we must choose a gauge for the photon field that would simplify the equations. In this work we employ the Coulomb gauge condition where $\partial_i\bs A^i=0$. In addition, from the first equation in system \eqref{system2} we can also choose $A^0=0$. Now by using the same method as shown in Ref. \cite{Ejlli:2016asd}, we expand the fields $A_i(\bs x, t)$ and $h_{ij}(\bs x, t)$ in the form
\begin{align}\label{field-expansion}
\bs A^i({\bs x}, t) &=\sum_{\lambda=x, y, z} e_\lambda^{i}(\hat{\bs n}) A_{\lambda}({\bs x, \omega})e^{-i \int \omega(t^\prime) dt^\prime},\quad  h_{ij}(\bs x, t)=\sum_{\lambda^\prime=\times, +} h_\lambda(\bs x, \omega) \textrm{e}_{ij}^{\lambda^\prime}(\hat{\bs n}) e^{-i \int \omega(t^\prime) dt^\prime},
\end{align}
where $e_\lambda^i$ is the photon polarization vector, $e_{ij}^{\lambda^\prime}$ is the GW polarization tensor with $\lambda^\prime$ indicating the polarization index or helicity state, and $\hat{\bs n}=\bs x/r$ with $r=|\bs x|$. Here $\hat{\bs n}$ is the direction of the propagation of the GW. Without any loss of generality, let us suppose now that the GW propagates in a given coordinate system along the $z$ axis, namely $\hat{\bs n}=\hat{\bs z}$. Since we are working in the Coulomb gauge where there is not a propagating longitudinal component for $\bs A^i$ and because $\bs x=r \hat{\bs z}$, we have that the third term on the left-hand side of the second equation in \eqref{system2}, namely $\partial^i\partial_\mu A^\mu$, is zero because of the Coulomb gauge and because $A^0=0$. In the equation governing the GW evolution (the third equation in \eqref{system2}), the last term $\bar B_i \bar B_j$, is a slowly varying function in space and time and can be neglected with respect to the interference terms $B_i \bar B_j$ and $\bar B_i B_j$.

Consider now the external magnetic field with components $\bar{\bs B}(\bs x, t)=[\bar B_x(\bs x, t), \bar B_y(\bs x, t), \bar B_z(\bs x, t)]$ and the vector potential with components $\bs A(\bs x, t)=[A_x(\bs x, t), A_y(\bs x, t), A_z(\bs x, t)]$. With the GW and electromagnetic wave propagating along the $\hat{\bs z}$ axis, $h_{ij}=h_{ij}(r, t), \bs A_i=\bs A_i(r, t)$ and with the field expansion \eqref{field-expansion}, the equations of motion \eqref{system2} for the GW tensor $h_{ij}$ in terms of the GW polarization states $h_+$ and $h_\times$ are given by
\begin{eqnarray}\label{GW-eq}
\left[\omega^2+\partial_r^2\right] h_+(r, \omega) &=& -\kappa \left[\partial_r A_x (r, \omega) \bar B_y+\partial_r A_y(r, \omega) \bar B_x\right],\nonumber\\
\left[\omega^2+\partial_r^2\right] h_\times (r, \omega) &=& \kappa \left[\partial_r A_x(r, \omega) \bar B_x-\partial_r A_y(r, \omega) \bar B_y\right],
\end{eqnarray}
where we used for the propagating electromagnetic wave $B_x (r, t)= - \partial_r A_y (r, t), B_y(r, t)=\partial_r A_x(r, t), B_z(r, t)=0$ with $\partial_r=\partial/\partial r$. In obtaining Eqs. \eqref{GW-eq} we used the fact that the GW polarization tensor is symmetric and depends only on $e_\lambda^{ij}(\hat{\bs z})$ and used the property $e_\lambda^{ij} e_{ij}^{\lambda^\prime}=2\delta_{\lambda\lambda^\prime}$.

In the case of equations of motion for the photon field $\bs A$ components in \eqref{system2}, we obtain
\begin{eqnarray}\label{PH-eq}
\left[ \omega^2 + \partial_r^2 - \Pi_{xx}(r, \omega) \right]A_x(r, \omega) - \Pi_{xy}(r, \omega)A_y(r, \omega) - \Pi_{xz}(r, \omega) A_z(\omega, r) &=&  \kappa \left[\partial_r h_+(r, \omega) \bar B_y-\partial_r h_\times (r, \omega) \bar B_x\right],\nonumber\\
\left[ \omega^2 + \partial_r^2 - \Pi_{yy}(r, \omega) \right]A_y (r, \omega) - \Pi_{yx}(r, \omega)A_x(r, \omega) - \Pi_{yz}(r, \omega) A_z(\omega, r) &=&  \kappa \left[\partial_r h_\times (r, \omega) \bar B_y + \partial_r h_+ (r, \omega) \bar B_x\right],\nonumber\\
\left[ \omega^2\delta_{zj} - \Pi_{z j}(r, \omega)\right] A_j(r, \omega) &=& 0,
\end{eqnarray}
where in the Coulomb gauge there is no propagating longitudinal electromagnetic wave $\partial_r A_z(r, t)=0$ and $\Pi^{ij}=\Pi_{ij}=\Pi_{ij}(r, \omega)$ are the elements of the photon polarization tensor calculated in the adiabatic limit $r^\prime\rightarrow r$. We may note that the third equation in the system \eqref{PH-eq} is actually a constraint on $A_z$. It can be shown \cite{Ejlli:2018ucq} that by solving this equation, namely by expressing $A_z$ in terms of the transverse photon states $A_x$ and $A_y$ and then substituting it in the first two equations in \eqref{PH-eq}, the components of $\Pi_{ij}$ for $i, j=x, y$ get a contribution from the longitudinal photon state. However, for the frequency range of the GWs and electromagnetic waves considered in this work, this extra contribution is very small and can safely be neglected.

The next step for solving Eqs. \eqref{GW-eq} and \eqref{PH-eq} is to look for solutions of field amplitudes of the form 
\begin{equation}\label{time-expansion}
h_{+, \times}(r, \omega)=\tilde{h} _{+, \times}(r, \omega)e^{i kr},\quad  A_{x, y}(r, \omega)=\tilde A_{x, y}(r, \omega)e^{i kr},
\end{equation}
where $k$ is the momentum of the fields corresponding to the mode $\bs k$. In addition, we work in the slowly varying envelope approximation (SVEA) which is a WKB-like approximation, namely that $|\partial_r \tilde h_{+, \times}|\ll | k \tilde h_{+, \times}|$ and $|\partial_r \tilde A_{x, y}|\ll |k \tilde A_{x, y}|$ with $(\omega^2 + \partial_r^2)(\cdot)=(\omega-i\partial_r)(\omega +i\partial_r)(\cdot)= (\omega+k)(\partial_t +i \partial_r)(\cdot)$. By using the expansion \eqref{time-expansion} in Eqs. \eqref{GW-eq} and \eqref{PH-eq} , we get the following system of first order differential equations for the field amplitudes $ h_{+, \times}$ and $A_{x, y}$
\begin{equation}\label{schr-eq}
(\omega + i\partial_r)\Psi(r, \omega)\bs I+M(r, \omega) \Psi(r, \omega)=0. 
\end{equation}
In \eqref{schr-eq} $\bs I$ is the unit matrix, $\Psi(r, \omega)=( h_\times, h_+,  A_x,  A_y)^\text{T}$ is a four component field, and $M(r, \omega)$ is the mass mixing matrix, which is given by
\begin{equation}\label{mixing-matrix}
 M=\begin{pmatrix}
  0 & 0 & -iM_{g\gamma}^x & iM_{g\gamma}^y \\
0 & 0 & iM_{g\gamma}^y & iM_{g\gamma}^x \\
iM_{g\gamma}^x & -iM_{g\gamma}^y & M_x & M_\text{CF} \\
-iM_{g\gamma}^y & -iM_{g\gamma}^x & M_\text{CF}^* & M_y
   \end{pmatrix},
   \end{equation}
where the elements of the mixing matrix $M$ are given by $M_{g\gamma}^{x}=\kappa\,k \bar B_x/(\omega+k)$, $M_{g\gamma}^y=\kappa \,k \bar B_{y}/(\omega+k)$, $M_x=-\Pi_{xx}/(\omega+k)$, and $M_y=-\Pi_{yy}/(\omega+k)$. Here $M_\text{CF}= - \Pi_{xy}/(\omega+ k)$ is a term that includes a combination of the CM effect and the Faraday effect and that depends on the magnetic field direction with respect to the photon propagation.  Here $\omega$ is the total energy of the fields, namely $\omega=\omega_\text{gr}=\omega_{\gamma}$. In this work all the particles participating in the mixing are assumed to be relativistic, namely $\omega+k\simeq 2k$.

 \section{Dispersive and coherence breaking effects in a magnetized plasma}
 \label{sec:3}

 In the previous section we have been able to reduce the equations of motion for the GRAPH mixing to a system of first order differential equations with variable coefficients. Before trying to look for a solution of the system \eqref{schr-eq} it is important to write the explicit expressions for $M_x, M_y$ ,and $M_\text{CF}$, which in turn depend on the elements of the photon polarization tensor in a magnetized medium. Here we present the explicit expressions for the elements $\Pi_{xx}, \Pi_{yy}$, and $\Pi_{xy}$ of the photon polarization tensor, and for a detailed discussion and derivation of these expressions see Refs. \cite{Ejlli:2016asd} and \cite{Ejlli:2018ucq}. The matrix elements $\Pi_{xx}$ and $\Pi_{yy}$ correspond to the modification of the dispersion and coherence breaking relations of the states $A_x$ and $A_y$, respectively; namely the momentum space Maxwell equations become, $\omega^2-k_{x, y}^2=\omega^2(1-n_{x, y}^2)=\Pi_{xx, yy}$, where $n_{x, y}$ are the total indexes of refraction. The expressions for the elements $\Pi_{xx}$ and $\Pi_{yy}$ are given \footnote{All expressions for the photon polarization tensor elements are derived under the conditions $\omega\neq \omega_c$ and $\omega>0$. In addition, propagating electromagnetic waves exist only when $\omega > \left(\pm \omega_c+\sqrt{\omega_c^2+4 \omega_\text{pl}^2}\right)/2$.} in Refs. \cite{Ejlli:2016asd} and \cite{Ejlli:2018ucq}
  \begin{equation}\label{pol-ele}
\Pi_{xx}=\frac{\omega^2 \omega_\text{pl}^2}{\omega^2-\omega_c^2} - \frac{\omega_\text{pl}^2\omega_c^2 \cos^2(\Theta)}{\omega^2-\omega_c^2}, \quad \Pi_{yy}=\frac{\omega^2\omega_\text{pl}^2}{\omega^2-\omega_c^2} - \frac{\omega_\text{pl}^2\omega_c^2 \sin^2(\Theta)\cos^2(\Phi)}{\omega^2-\omega_c^2},
\end{equation}
where $\omega_\text{pl}=\sqrt{4\pi\alpha n_e/m_e}$ is the plasma frequency and $\omega_c=e \bar B/m_e$ is the cyclotron frequency. Here $m_e$ is the electron mass, $e$ is the electron charge, $n_e$ is the number density of the free electrons in the plasma and $\bar B(\bs x, t)=|\bar{\bs B}(\bs x, t)|$ is the external magnetic field strength. In addition, $\Theta$ is the polar angle of the external magnetic field with respect to the $x$ axis which points to the North and $\Phi$ is the azimuthal angle of the external magnetic field with respect to the $y$ axis which points outward. For this configuration, we can write $\bar{\bs B}(\bs x, t)=[\bar B_x (\bs x, t)), \bar B_y (\bs x, t), \bar B_z (\bs x, t)]=\bar B (\bs x, t)\left[ \cos(\Theta), \sin(\Theta)\cos(\Phi), \sin(\Theta)\sin(\Phi)\right]$.

The firsts terms in $\Pi_{xx}$ and $\Pi_{yy}$ in \eqref{pol-ele} correspond to the effect of only electronic plasma to the polarization tensor. The second terms in \eqref{pol-ele} correspond to the CM effect in plasma since this effect is proportional to $\bar B^2$ (see Fig. \ref{Graphax}). On the other hand, the element $\Pi_{xy}$ is given by
\begin{equation}\label{CM-pol}
\Pi_{xy}=-\frac{\omega_\text{pl}^2\,\omega_c^2 \sin(2 \Theta)\cos(\Phi)}{2\,(\omega^2-\omega_c^2)} - i \frac{\omega_\text{pl}^2\omega_c\,\omega\, \sin(\Theta)\sin(\Phi)}{\omega^2-\omega_c^2}.
\end{equation}
The first term in \eqref{CM-pol} is due to the CM effect while the second term corresponds to the Faraday effect in plasma. Since the second term is imaginary, it essentially means that the Faraday effect changes the intensity of each photon polarization state, namely a coherence breaking effect. Typically in the literature it is used to get rid of the first term in $\Pi_{xy}$ by choosing $\Phi=\pi/2$, namely by choosing the external magnetic field $\bar{\bs B}$ and the photon wave vector $\bs k$ in the $xz$ plane. In such a case $\Pi_{xy}$ is purely imaginary and it includes the Faraday effect only.

In many situations one can simplify the expressions of the elements of the photon polarization tensor by making some reasonable assumptions on the magnitude of the photon frequency with respect to the plasma and cyclotron frequencies. The numerical value of the angular plasma frequency can be written as $\omega_\text{pl}=5.64\times 10^4 \sqrt{n_e/\text{cm}^3}$ (rad/s) or $\nu_\text{pl}=\omega_\text{pl}/(2\pi)=8976.33 \sqrt{n_e/\text{cm}^3}$ (Hz) for the frequency. On the other hand, the numerical value of the cyclotron angular frequency is given by $\omega_c=1.76\times 10^7(\bar B/\text{G})$ (rad/s). The cases when $\omega\gg \omega_\text{pl}$ and  $\omega\gg \omega_c$ are of particular interest in many situations and especially in this work. As shown in the previous section, the quantities $\omega_c$ and $\omega_\text{pl}$ do not explicitly depend on the time $t$ but do explicitly depend on the distance $r$. However, in the case of photon propagation in an expanding universe, we can express the distance $r$ in terms of the cosmological time $t$ as $r=r(t)$. Consequently, each quantity that explicitly depends on $r$, also implicitly depends on $t$ because of $r=r(t)$. Therefore, the conditions $\omega\gg \omega_\text{pl}$ and  $\omega\gg \omega_c$, in an expanding universe, are, respectively, satisfied when
\begin{equation}\label{nu-con-1}
\left(\frac{\nu_0}{\text{Hz}}\right)\gg 8976.33 \left(\frac{0.76\,n_B(t_0) X_e(t)}{\text{cm}^3}\right)^{1/2} \left(\frac{a(t_0)}{a(t)}\right)^{1/2} \quad \text{and} \quad \left(\frac{\nu_0}{\text{Hz}}\right)\gg 2.8\times 10^6 \left(\frac{\bar B_0}{\text{G}}\right)\left(\frac{a(t_0)}{a(t)}\right),
\end{equation}
where we expressed $\nu(t)=\nu_0 [a(t_0)/a(t)]$ with $\nu_0$ being the frequency of the electromagnetic radiation at the present time $t=t_0$ and with $a(t)$ being the universe expansion scale factor, and $\bar B_0=\bar B(t_0)$ is the magnetic field strength at the present time \footnote{In what follows we assume that the magnetic field amplitude depends only on time $t$ and not on $\bs x$.}. Here we expressed the number density of free electrons as $n_e(t)\simeq 0.76\, n_B(t_0) X_e(t)[a(t_0)/a(t)]^3$ where $n_B(t_0)$ is the total baryon number density at the present time and $X_e(t)$ is the ionization fraction of the free electrons. The factor of $0.76$ takes into account the contribution of hydrogen atoms to the free electrons at the post-decoupling time. 

By taking, for example $n_B(t_0)\simeq 2.47\times 10^{-7}$ cm$^{-3}$ ,as given by the Planck Collaboration \cite{Ade:2015xua} and expressing $a(t_0)/a(t)=1+z$ where $z$ is the source redshift, we can write the conditions \eqref{nu-con-1} as
\begin{equation}\label{nu-con-2}
\left(\frac{\nu_0}{\text{Hz}}\right)\gg 3.88 \left(1+z \right)^{1/2} \quad \text{and} \quad \left(\frac{\nu_0}{\text{Hz}}\right)\gg 2.8 \times 10^6 \left(\frac{\bar B_0}{\text{G}}\right) \left( 1+z \right),
\end{equation}
where at the post-decoupling epoch we can safely assume $X_e(t)\simeq 1$. In most situations, photon frequencies that satisfy the first condition in \eqref{nu-con-2}, also satisfy the second condition in \eqref{nu-con-2} for realistic values of $\bar B_0$ and for redshifts $z\lesssim 20$. After these considerations, we can approximate the expressions of the elements of the photon polarization tensor as
\begin{align}\label{pol-ele-1}
\Pi_{xx} & \simeq \omega_\text{pl}^2\left[1 - \frac{\omega_c^2 \cos^2(\Theta)}{\omega^2}\right], \quad \Pi_{yy} \simeq \omega_\text{pl}^2\left[1- \frac{\omega_c^2 \sin^2(\Theta)\cos^2(\Phi)}{\omega^2}\right],\nonumber \\  \Pi_{xy} & \simeq -\frac{\omega_\text{pl}^2\,\omega_c^2 \sin(2 \Theta)\cos(\Phi)}{2\, \omega^2} - i \frac{\omega_\text{pl}^2\omega_c \sin(\Theta)\sin(\Phi)}{\omega}.
\end{align}
There is another fact about the expressions in \eqref{pol-ele-1} that is important to mention now. The second terms in $\Pi_{xx, yy}$, which essentially correspond to the CM effect, are indeed very small quantities with respect to unity in the case $\omega\gg \omega_c, \omega_\text{pl}$ and can be neglected in many cases. The only case when these quantities cannot be neglected is when we have to deal with the difference $\Pi_{xx}-\Pi_{yy}$ or vice versa. Regarding the term $\Pi_{xy}$, we may note that in the cases when $\sin(\Theta)\sin(\Phi)\neq 0$, the magnitude of the imaginary term that essentially corresponds to the Faraday effect is much bigger than the magnitude of the real term that corresponds to the CM effect.

 \section{Perturbative solutions of the equations of motion}
 \label{sec:4}

In this section we focus on perturbative solutions of the equations of motion \eqref{schr-eq}. The main reason to look for such solutions is because they do not exist for exact closed solutions except in some particular cases which are of no interest in this work. Here we employ a similar formalism as in quantum mechanics, namely similar to the time dependent perturbation theory, where usually one writes the total Hamiltonian of the system as the sum of a ``free'' term plus a time dependent small interaction term. In our specific case the mass mixing matrix $M$ plays the role of the total Hamiltonian and which depends on the distance rather than the time. Consequently, in our case we may split the mass mixing matrix in the following way $M(\omega, r)=M_0(\omega, r)+M_1(\omega, r)$ where $M_0(\omega, r)$ is a matrix which would enter the equations of motion \eqref{schr-eq} in the case when GWs would not be present and $M_1(\omega, r)$ is a perturbation matrix that takes into account the interaction of GWs with the external magnetic field
\begin{equation}
M_0(\omega, r)=\begin{pmatrix}
0 & 0 & 0 & 0 \\
0 & 0 & 0 & 0 \\
0 & 0 & M_x & M_\text{CF} \\
0 & 0 & M_\text{CF}^* & M_y
   \end{pmatrix}, \quad
M_1(\omega, r)=\begin{pmatrix}
0 & 0 & -iM_{g\gamma}^x & iM_{g\gamma}^y \\
0 & 0 & iM_{g\gamma}^y & iM_{g\gamma}^x \\
iM_{g\gamma}^x & -iM_{g\gamma}^y & 0 & 0 \\
-iM_{g\gamma}^y & -iM_{g\gamma}^x & 0 & 0
   \end{pmatrix}.
      \end{equation}

In the case where GWs are missing, the matrix $M_0$ would enter Eq. \eqref{schr-eq} in the form $\left(\omega + i\partial_r\right)\Psi(\omega, r)\bs I+M_0(\omega, r)\Psi(\omega, r)=0$ without the presence of the perturbation matrix $M_1$. However, even in the absence of the perturbation matrix $M_1$, it is not possible to find a closed analytical solution for Eq. \eqref{schr-eq} since we are dealing with a first order system of differential equations with variable coefficients with analytic solutions are rare except in some particular cases. There is a possibility to solve analytically Eq. \eqref{schr-eq} for $M=M_0$ in the case when $M_x=M_y$. In fact, we may note from the expressions of $\Pi_{xx, yy}$ in \eqref{pol-ele-1} that in the case when $\omega \gg \omega_c$, the CM effect can be neglected with respect to the plasma effect. In this regime we may approximate $M_x\simeq M_y$ in $M_0$. In this case the commutator $[M_0(\omega, r), M_0(\omega, r^\prime)]=0$ and the solution of Eq. \eqref{schr-eq} for $M=M_0$ is given by $\Psi(\omega, r)=U(r, r_i)\Psi(\omega, r_i)$ where $U$ is the usual unitary evolution operator which is given by $U(r, r_i)=\exp[-i\int_{r_i}^r dr^\prime \left(-\omega(r^\prime)\bs I- M_0(r^\prime)\right)]$.

In the case when the interaction is present, namely when $M=M_0+M_1$, in order to solve Eq. \eqref{schr-eq}, it is convenient to move to the ``interaction picture'' by defining $\Psi_\text{int}(\omega, r)=U^\dagger(r, r_i) \Psi(\omega, r)$ and $M_\text{int}(\omega, r)=U^\dagger(r, r_i) M_1(\omega, r) U(r, r_i)$. In the ``interaction picture'', Eq. \eqref{schr-eq} becomes $i\partial_r\Psi_\text{int}(\omega, r)=M_\text{int}(\omega, r)\Psi_\text{int}(\omega, r)$. By using an iterative procedure, we find the following perturbative solution for $\Psi_\text{int}(\omega, r)$ to first and second orders in the perturbation matrix $M_\text{int}(\omega, r)$
\begin{equation}
\Psi_\text{int}^{(1)}(\omega, r)=-i\int_{r_i}^r dr^\prime\, M_\text{int}(\omega, r^\prime)\Psi(r_i, \omega_i), \quad \Psi_\text{int}^{(2)}(\omega, r)=-\int_{r_i}^r \int_{r_i}^{r^\prime} dr^\prime\, dr^{\prime\prime}\,  M_\text{int}(\omega, r^\prime)\,M_\text{int}(\omega, r^{\prime\prime})\Psi(r_i, \omega_i),
\end{equation}
where $\Psi_\text{int}^{(0)}(\omega, r)=\Psi(\omega_i, r_i)$ and $\Psi_\text{int}(\omega, r)=\Psi_\text{int}^{(0)}(\omega, r)+\Psi_\text{int}^{(1)}(\omega, r)+\Psi_\text{int}^{(2)}(\omega, r)+ \text{higher order terms}$. Since we have that the elements $|\int_{r_i}^r dr^\prime M_{1, ij}(r^\prime)|\ll 1$ for reasonable values of the parameters, the series expansion converges rapidly, and consequently it is not necessary to go beyond the first order expansion. Therefore, by performing several operations and by dropping for the moment the dependence of the fields on $\omega$, we get the following solutions for the field amplitudes in the interaction picture up to the first order in perturbation theory: 
\begin{align}\label{eq-solutions}
h_\times(r) &= h_\times(r_i) - A_x(r_i)\,\int_{r_i}^{r} dr^{\prime}\left( \cos\left[\sqrt{\mathcal M_\text{CF}(r^\prime)}\sqrt{\mathcal M_\text{CF}^*(r^\prime)}\right] M_{g\gamma}^x(r^\prime) - i\, \mathcal C(r^\prime)\sin\left[\sqrt{\mathcal M_\text{CF}(r^\prime)}\sqrt{\mathcal M_\text{CF}^*(r^\prime)}\right] \right. \nonumber\\ & \left. \times  M_{g\gamma}^y(r^\prime) \right) e^{i  M_1(r^\prime)} +  A_y(r_i)\,\int_{r_i}^{r} dr^{\prime}\left( \cos\left[\sqrt{\mathcal M_\text{CF}(r^\prime)}\sqrt{\mathcal M_\text{CF}^*(r^\prime)}\right] M_{g\gamma}^y(r^\prime) - i\, \mathcal C^{-1}(r^\prime)\times \right. \nonumber\\ & \left.  \sin\left[\sqrt{\mathcal M_\text{CF}(r^\prime)}\sqrt{\mathcal M_\text{CF}^*(r^\prime)}\right] M_{g\gamma}^x(r^\prime) \right) e^{i  M_1(r^\prime)},\nonumber
\end{align}
\begin{align}
h_+(r) &= h_+(r_i) + A_x(r_i)\,\int_{r_i}^{r} dr^{\prime}\left( \cos\left[\sqrt{\mathcal M_\text{CF}(r^\prime)}\sqrt{\mathcal M_\text{CF}^*(r^\prime)}\right] M_{g\gamma}^y(r^\prime) + i\, \mathcal C(r^\prime)\sin\left[\sqrt{\mathcal M_\text{CF}(r^\prime)}\sqrt{\mathcal M_\text{CF}^*(r^\prime)}\right] \right. \nonumber\\ & \left. \times  M_{g\gamma}^x(r^\prime) \right) e^{i  M_1(r^\prime)} +  A_y(r_i)\,\int_{r_i}^{r} dr^{\prime}\left( \cos\left[\sqrt{\mathcal M_\text{CF}(r^\prime)}\sqrt{\mathcal M_\text{CF}^*(r^\prime)}\right] M_{g\gamma}^x(r^\prime) + i\, \mathcal C^{-1}(r^\prime)\times \right. \nonumber\\ & \left.  \sin\left[\sqrt{\mathcal M_\text{CF}(r^\prime)}\sqrt{\mathcal M_\text{CF}^*(r^\prime)}\right] M_{g\gamma}^y(r^\prime) \right) e^{i  M_1(r^\prime)},\nonumber
\end{align}
\begin{align}
A_x(r) &= A_x(r_i) + h_\times(r_i)\,\int_{r_i}^{r} dr^{\prime}\left( \cos\left[\sqrt{\mathcal M_\text{CF}(r^\prime)}\sqrt{\mathcal M_\text{CF}^*(r^\prime)}\right] M_{g\gamma}^x(r^\prime) + i\, \mathcal C^{-1}(r^\prime)\sin\left[\sqrt{\mathcal M_\text{CF}(r^\prime)}\sqrt{\mathcal M_\text{CF}^*(r^\prime)}\right] \right. \nonumber\\ & \left. \times  M_{g\gamma}^y(r^\prime) \right) e^{- i  M_1(r^\prime)} -  h_+(r_i)\,\int_{r_i}^{r} dr^{\prime}\left( \cos\left[\sqrt{\mathcal M_\text{CF}(r^\prime)}\sqrt{\mathcal M_\text{CF}^*(r^\prime)}\right] M_{g\gamma}^y(r^\prime) - i\, \mathcal C^{-1}(r^\prime)\times \right. \nonumber\\ & \left.  \sin\left[\sqrt{\mathcal M_\text{CF}(r^\prime)}\sqrt{\mathcal M_\text{CF}^*(r^\prime)}\right] M_{g\gamma}^x(r^\prime) \right) e^{- i  M_1(r^\prime)},\nonumber\\
A_y(r) &= A_y(r_i) - h_\times(r_i)\,\int_{r_i}^{r} dr^{\prime}\left( \cos\left[\sqrt{\mathcal M_\text{CF}(r^\prime)}\sqrt{\mathcal M_\text{CF}^*(r^\prime)}\right] M_{g\gamma}^y(r^\prime) + i\, \mathcal C(r^\prime)\sin\left[\sqrt{\mathcal M_\text{CF}(r^\prime)}\sqrt{\mathcal M_\text{CF}^*(r^\prime)}\right] \right. \nonumber\\ & \left. \times  M_{g\gamma}^x(r^\prime) \right) e^{- i  M_1(r^\prime)} -  h_+(r_i)\,\int_{r_i}^{r} dr^{\prime}\left( \cos\left[\sqrt{\mathcal M_\text{CF}(r^\prime)}\sqrt{\mathcal M_\text{CF}^*(r^\prime)}\right] M_{g\gamma}^x(r^\prime) - i\, \mathcal C(r^\prime)\times \right. \nonumber\\ & \left.  \sin\left[\sqrt{\mathcal M_\text{CF}(r^\prime)}\sqrt{\mathcal M_\text{CF}^*(r^\prime)}\right] M_{g\gamma}^y(r^\prime) \right) e^{- i  M_1(r^\prime)},
\end{align}
where we have defined 
\begin{eqnarray*}
M_{\{1, 2\}}(r) &\equiv& \int_{r_i}^r dr^\prime\, M_{\{x, y\}}(r^\prime),\quad
\mathcal M_\text{CF}(r)  \equiv   \int_{r_i}^r dr^\prime M_\text{CF}(r^\prime), \\ \mathcal M_\text{CF}^*(r) & \equiv &   \int_{r_i}^r dr^\prime M_\text{CF}^*(r^\prime), \quad \mathcal C(r) \equiv \sqrt{\mathcal M_\text{CF}^*(r)/\mathcal M_\text{CF}(r)},
\end{eqnarray*}
with $r_i$ being the initial distance and for simplicity in \eqref{eq-solutions} we dropped the subscript ''int'' of the interaction picture field amplitudes. In obtaining the solutions \eqref{eq-solutions}, we have assumed that $\mathcal M_\text{CF}^*(r)\neq 0$ and $\mathcal M_\text{CF}(r)\neq 0$. In addition, since the gravitons are assumed to be exactly massless, we have that $M_{\times, +}=0$. As already mentioned above, on obtaining the solutions \eqref{eq-solutions} we have assumed that $M_y\simeq M_x$ and therefore we have approximated $M_2\simeq M_1$. We may also note from the solutions \eqref{eq-solutions} that in the expressions of $h_{\times, +}(r)$,$h_{+, \times}(r_i)$ do not appear; namely there is no mixing between the states $h_{\times, +}$ at first order in the perturbation theory. Such a mixing appears starting from the second order of iteration. Analog conclusions apply also for the photon states $A_{x, y}$.

As it will be clear in what follows, it is very convenient in many calculations involving the photon amplitudes to write  
\begin{align}\label{ph-amplitude}
 A_x(r) & = I_1(r)  h_\times( r_i)-I_2(r) h_+( r_i)+ A_x(r_i),\nonumber\\
 A_y(r) & =-I_3(r) h_\times(r_i)-I_4(r) h_+( r_i) + A_y(r_i),
\end{align}
where we have defined
\begin{align}\label{integrals}
I_1(r) &\equiv \int_{r_i}^{r} dr^{\prime}\left( \cos\left[\sqrt{\mathcal M_\text{CF}(r^\prime)}\sqrt{\mathcal M_\text{CF}^*(r^\prime)}\right] M_{g\gamma}^x(r^\prime) + i\, \mathcal C^{-1}(r^\prime)\sin\left[\sqrt{\mathcal M_\text{CF}(r^\prime)}\sqrt{\mathcal M_\text{CF}^*(r^\prime)}\right]  M_{g\gamma}^y(r^\prime) \right) e^{- i  M_1(r^\prime)},\nonumber \\
I_2(r) &\equiv \int_{r_i}^{r} dr^{\prime}\left( \cos\left[\sqrt{\mathcal M_\text{CF}(r^\prime)}\sqrt{\mathcal M_\text{CF}^*(r^\prime)}\right] M_{g\gamma}^y(r^\prime) - i\, \mathcal C^{-1}(r^\prime) \sin\left[\sqrt{\mathcal M_\text{CF}(r^\prime)}\sqrt{\mathcal M_\text{CF}^*(r^\prime)}\right] M_{g\gamma}^x(r^\prime) \right) e^{- i  M_1(r^\prime)},\nonumber \\
I_3(r) &\equiv  \int_{r_i}^{r} dr^{\prime}\left( \cos\left[\sqrt{\mathcal M_\text{CF}(r^\prime)}\sqrt{\mathcal M_\text{CF}^*(r^\prime)}\right] M_{g\gamma}^y(r^\prime) + i\, \mathcal C(r^\prime)\sin\left[\sqrt{\mathcal M_\text{CF}(r^\prime)}\sqrt{\mathcal M_\text{CF}^*(r^\prime)}\right]  M_{g\gamma}^x(r^\prime) \right) e^{- i  M_1(r^\prime)},\nonumber \\
I_4(r) & \equiv \int_{r_i}^{r} dr^{\prime}\left( \cos\left[\sqrt{\mathcal M_\text{CF}(r^\prime)}\sqrt{\mathcal M_\text{CF}^*(r^\prime)}\right] M_{g\gamma}^x(r^\prime) - i\, \mathcal C(r^\prime)  \sin\left[\sqrt{\mathcal M_\text{CF}(r^\prime)}\sqrt{\mathcal M_\text{CF}^*(r^\prime)}\right] M_{g\gamma}^y(r^\prime) \right) e^{- i  M_1(r^\prime)}.
\end{align}

\section{Generation of electromagnetic radiation and Stokes parameters}
\label{sec:5}

In this section we focus our attention on the  generation of the electromagnetic radiation for the GRAPH mixing. In particular, here we consider the situation of a source that emits GWs, and we want to calculate useful quantities regarding the electromagnetic radiation such as the intensity and the power. To have a full picture of the generated electromagnetic radiation in the GRAPH mixing, it is quite convenient to start with the Stokes parameters that give a complete description of the intensity and polarization state of the electromagnetic radiation. They are usually defined in terms of the transverse electric field amplitudes $E_x$ and $E_y$ ($\bs E(\bs x, t)= [E_x(\bs x, t), E_y(\bs x, t)]$) at a fixed point in space $\bs x$ as 
\begin{eqnarray}\label{Stokes-par}
I_\gamma(\bs x, t) & \equiv & |E_x(\bs x, t)|^2+|E_y(\bs x, t)|^2, \quad Q(\bs x, t) \equiv |E_x(\bs x, t)|^2-|E_y(\bs x, t)|^2,\nonumber\\
 U(\bs x, t) & \equiv & 2\,\text{Re}\left\{E_x(\bs x, t)E_y^*(\bs x, t) \right\}, \quad  V(\bs x, t) \equiv -2\,\text{Im}\left\{E_x(\bs x, t)E_y^*(\bs x, t) \right\}.
 \end{eqnarray}

Consider now the situation where a given source emits GWs with polarization states $ h_{\times, +}$ and initially photons are not present. By reintroducing the dependence of the fields on $\omega$ again, the amplitudes of the photon states $ A_{x, y}$ at the distance $r$ from the source, given in expression \eqref{ph-amplitude}, can be written as
\begin{eqnarray}\label{ph-amplitude-1}
 A_x(r, \omega) &=&I_1(r) h_\times(r_i, \omega_i)-I_2(r) h_+(r_i, \omega_i),\nonumber \\
 A_y(r, \omega) &=&-I_3(r) h_\times(r_i, \omega_i)-I_4( r) h_+(r_i, \omega_i),
\end{eqnarray}
where the dependence of the fields on $\omega$ appears through the integrals $I_{1, 2, 3, 4}$ which do depend on $\omega$ parametrically, $I_{1, 2, 3, 4}(r; \omega)$.
Let us concentrate on the calculation of the photon intensity $I_\gamma(r, t)$ and other Stokes parameters. In this case we need the explicit expressions for the electric field amplitudes $E_x$ and $E_y$ which are, respectively, given by $E_x(\bs x, t)=-\partial_t A_x(\bs x, t)-\nabla\cdot A^0(\bs x, t)$ and $E_y(\bs x, t)=-\partial_t A_y(\bs x, t)-\nabla\cdot A^0(\bs x, t)$. If the generated electromagnetic wave travels along the $z$ axis, then we have at the distance $r$ from the source that $A_{x, y}(r, t)= A_{x, y}(r, \omega)\,e^{-i\int \omega(t^\prime) dt^\prime}$. On the other hand, the expression for the scalar potential, $A^0(\bs x, t)=0$ by choice. After these considerations, we can write the expressions for the components of the electric field in the SVEA approximation, for an electromagnetic wave propagating along the $z$ axis at a distance $r$ from the source
\begin{equation}\label{ele-field}
E_{x, y}(r, t) \simeq -i\,\omega(t)\, A_{x, y}(r, \omega)\,e^{-i\int \omega(t^\prime) dt^\prime}=-i \omega(t)\, A_{x, y}(r, t).
\end{equation}

With the expression for the electric field components given in \eqref{ele-field}, we can easily calculate the expression for the Stokes parameters for the generated electromagnetic field radiation, which are given by
\begin{eqnarray}\label{ph-int}
I_\gamma(r, t) &=& \omega^2(t)\left[| A_x(r, t)|^2 + | A_y(r, t)|^2 \right], \quad Q(r, t)=\omega^2(t)\left[| A_x(r, t)|^2 - | A_y(r, t)|^2 \right],\nonumber\\
U(r, t) &=& 2\,\omega^2(t)\,\text{Re}\left\{ A_x(r, t) A_y^*(r, t)\right\}, \quad V(r, t) =- 2\,\omega^2(t)\,\text{Im}\left\{ A_x(r, t) A_y^*(r, t)\right\}.
\end{eqnarray}
Now by using the expressions \eqref{ph-amplitude-1} in \eqref{ph-int}, we get
\begin{align}\label{ph-int-1}
I_\gamma(r, t) &= \omega^2\left[\left(|I_1(r)|^2+|I_3(r)|^2\right)| h_\times(r_i)|^2 +\left(|I_2(r)|^2+|I_4(r)|^2\right)| h_+(r_i)|^2  \right. \nonumber\\ & \left. + 2\, \text{Re}\left\{\left[I_3(r)I_4^*(r)-I_1(r)I_2^*(r)\right] h_\times(r_i) h_+^*(r_i)\right\}\right],\nonumber \\
Q(r, t) &= \omega^2\left[\left(|I_1(r)|^2 - |I_3(r)|^2\right)| h_\times(r_i)|^2 +\left(|I_2(r)|^2 - |I_4(r)|^2\right)| h_+(r_i)|^2 \right. \nonumber\\ & \left. - 2\, \text{Re}\left\{\left[I_1(r)I_2^*(r) + I_3(r)I_4^*(r)\right] h_\times(r_i) h_+^*(r_i)\right\}\right],\nonumber\\
U(r, t) &= 2\,\omega^2\,\text{Re}\left\{-I_1(r)I_3^*(r)| h_\times(r_i)|^2 + I_2(r)I_4^*(r)| h_+(r_i)|^2 -I_1(r)I_4^*(r) h_\times(r_i) h_+^*(r_i) + I_2(r)I_3^*(r) h_+(r_i) h_\times^*(r_i)\right\},\nonumber \\
V(r, t) &= 2\,\omega^2\,\text{Im}\left\{I_1(r)I_3^*(r)| h_\times(r_i)|^2 - I_2(r)I_4^*(r)| h_+(r_i)|^2 +I_1(r)I_4^*(r) h_\times(r_i) h_+^*(r_i) - I_2(r)I_3^*(r) h_+(r_i) h_\times^*(r_i)\right\}.
\end{align}
The expressions for the Stokes parameters given in \eqref{ph-int-1} are one of the most important results in this work and will be the basis of our study of the generation of the electromagnetic radiation in the GRAPH mixing. One should keep in mind that expressions \eqref{ph-int-1} have been obtained by using expressions \eqref{eq-solutions} for the field amplitudes in the interaction picture. By carefully going back to the ordinary picture $\Psi(\omega, r)=U(r, r_i) \Psi_\text{int}(\omega, r)$, one can easily check that the intensity Stokes parameter in \eqref{ph-int-1} is invariant under the field transformation $\Psi(\omega, r)=U(r, r_i) \Psi_\text{int}(\omega, r)$ while expressions for other Stokes parameters $Q, U$ and $V$ change due to a contribution of the Faraday and CM effects.

\section{Evaluation of the integrals, $I_1, I_2, I_3,$ and $I_4$}
\label{sec:6}

As we see from the expressions of the Stokes parameters given in \eqref{ph-int-1}, in order to calculate them, first we must calculate the integrals $I_1, I_2, I_3$, and $I_3$ which do appear in each of the parameters. The explicit expressions for the integrals $I_1, I_2, I_3$, and $I_4$ are given in \eqref{integrals}. We may note that each of them contains to first order in perturbation theory, the integration over the distance of either $M_{g\gamma}^{x}$ or $M_{g\gamma}^y$ times trigonometric functions containing the CM and Faraday effects and also the exponential of plasma effects. Before evaluating the integrals, we must explicitly write all quantities tha enter in each of them. 

The explicit expressions for $M_1, M_2$, and $M_\text{CF}$, for $\omega\simeq k$, are given by
\begin{align}\label{matr-ele}
M_1(r) &=\int_{r_i}^r dr^\prime M_x(r^\prime)=-\int_{r_i}^r dr^\prime \left(\frac{\Pi_{xx}}{2\omega}\right) \simeq -\int_{r_i}^r dr^\prime\, \frac{\omega_\text{pl}^2}{2 \omega}\left[1 - \frac{\omega_c^2 \cos^2(\Theta)}{\omega^2}\right],\nonumber\\
M_2(r) &=\int_{r_i}^r dr^\prime M_y(r^\prime)=-\int_{r_i}^r dr^\prime \left(\frac{\Pi_{yy}}{2\omega}\right) \simeq -\int_{r_i}^r dr^\prime\, \frac{\omega_\text{pl}^2}{2 \omega}\left[1 - \frac{\omega_c^2 \sin^2(\Theta)\cos^2(\Phi)}{\omega^2}\right],\nonumber\\
M_\text{CF}(r) &= M_\text{C}(r)+i M_\text{F}(r)=-\frac{\Pi_{xy}}{2\omega} \simeq \frac{\omega_\text{pl}^2\,\omega_c^2 \sin(2 \Theta)\cos(\Phi)}{4\, \omega^3} + i \frac{\omega_\text{pl}^2\omega_c \sin(\Theta)\sin(\Phi)}{2\,\omega^2},
\end{align}
where we used the expressions for the elements of the photon polarization tensor given in \eqref{pol-ele}. 
On the other hand, the explicit expressions for $M_{g\gamma}^x$ and $M_{g\gamma}^y$ are, respectively, given by $M_{g\gamma}^x(r)=\kappa \bar B \cos(\Theta)/2$ and $M_{g\gamma}^y(r)=\kappa \bar B \sin(\Theta)\cos(\Phi)/2$. The quantities $\omega_\text{pl}, \omega_c, \omega$, and $\bar B$ in \eqref{matr-ele} depend on the distance $r$ and implicitly depend on the time in an expanding universe; see below. Also the angles $\Theta$ and $\Phi$ may depend on the time but in this work we assume that the external magnetic field direction at a given point $\bs x$ does not change in time, therefore $\Theta$ and $\Phi$ are assumed to be constant in time. In \eqref{matr-ele} we have expanded $M_\text{CF}=M_\text{C}+iM_\text{F}$ with $M_\text{C}$ being the term corresponding to the CM effect and $M_\text{F}$ being the term corresponding to the Faraday effect.

After the considerations made above, let us now focus on the calculations of the integrals $I_{1, 2}$ and $I_{3, 4}$. As we may note, the integrals $I_1$ and $I_2$ have the same structure, and therefore it will be sufficient to calculate only one of them. 
At this stage it is more useful to express each space dependent quantity as a function of the redshift $z$ since we are going to deal with electromagnetic radiation and GWs propagating in an expanding universe. For relativistic particles propagating in null geodesics we have that the line element $ds^2=0$ which implies that $dt=dr$ where $r$ is the light traveled distance and $t$ is the cosmological time. In this case the integration over the distance in each integral is replaced with the integration over the redshift $z$ by using the following prescription
\begin{equation}\label{light-t-distance}
\int_{r_i}^r\, dr^\prime (...)=\int_{t_i}^t\, dt^\prime (...)=\int_{z}^{z_i}\,\frac{dz^{\prime}}{H_0(1+z^\prime)\sqrt{\Omega_\Lambda+\Omega_\text{M}(1+z^\prime)^3+\Omega_\text{R}(1+z^\prime)^4}} (...),
\end{equation}
where $\Omega_\Lambda\simeq 0.68$ is the present epoch density parameter of the vacuum energy, $\Omega_\text{M}\simeq 0.31$ is the present epoch density parameter of the nonrelativistic matter and $\Omega_\text{R}\ll 1$ is the present epoch density parameter of the relativistic matter that essentially includes relativistic photons and neutrinos. Here we are assuming a universe with zero spatial curvature, namely $\Omega_\kappa=0$. In addition $r_i<r$ and $z<z_i$ where $z_i$ is the redshift of the GWs emitting source. In general for astrophysical sources of GWs that are located at relatively low redshifts, one can safely neglect the contribution of the relativistic matter to the total energy density. Moreover, in many cases it is quite accurate to approximate, $\sqrt{\Omega_\Lambda +\Omega_\text{M}(1+z)^3}\simeq \sqrt{\Omega_\Lambda}$ for $z\ll \left[(\Omega_\Lambda/\Omega_\textrm{M})^{1/3}-1\right]\simeq 0.29$ or  $\sqrt{\Omega_\Lambda +\Omega_\text{M}(1+z)^3}\simeq \sqrt{\Omega_\textrm{M}(1+z)^3}$ for $z\gg 0.29$.

In a case when $\omega\gg \omega_c$, we may neglect the second terms proportional to the plasma frequency in $M_{1, 2}$ in \eqref{matr-ele} and approximate
\begin{align}\label{M1}
M_1(z) & \simeq M_2(z)=-\int_{z}^{z_i}\,\frac{dz^{\prime}}{H_0(1+z^\prime)\sqrt{\Omega_\Lambda+\Omega_\text{M}(1+z^\prime)^3}}\left(\frac{\omega_\textrm{pl}^2(z^\prime)}{2\,\omega(z^\prime)}\right), \nonumber \\ & \simeq \begin{cases} -2\, \mathcal A_1\,\Omega_\text{M}^{-1/2}\,H_0^{-1}\,\left(\sqrt{1+z_i}-\sqrt{1+z}\right),\quad \text{for}\quad z\gg \left[(\Omega_\Lambda/\Omega_\textrm{M})^{1/3}-1\right], \\
- (\mathcal A_1/2)\,\Omega_\Lambda^{-1/2}\,H_0^{-1}\,\left[(z_i^2-z^2)+2\,(z_i-z)\right],\quad \text{for}\quad z\ll \left[(\Omega_\Lambda/\Omega_\textrm{M})^{1/3}-1\right],
\end{cases}
\end{align}
where we expressed the plasma and incident photon frequencies as a function of the redshift as shown in Sec. \ref{sec:3}, namely $\omega_\text{pl}^2/(2\omega)=\mathcal A_1 (1+z)^2$ where $\mathcal A_1\equiv 3.12\times 10^{-14}(\text{Hz}/\nu_0)$ (eV). Since in this work we focus on the post-decoupling epoch, we assume that $X_e(z)\simeq 1$. Now in order to calculate the integrals in \eqref{integrals}, let us write the amplitude of the external magnetic field as $\bar B(z)=\bar B_0(1+z)^2$, which is derived from the assumption that the magnetic flux in the cosmological plasma is a conserved quantity. The other quantities that we be useful in what follows are $\mathcal M_\text{C}$ and $\mathcal M_\text{F}$. The expression for $\mathcal M_{F}$ can be calculated exactly\footnote{The expression for $\mathcal M_\text{F}$ is exact for fixed values of $\Omega_{\Lambda, \text{M}}$ that are found experimentally by Planck Collaboration. However, the general expression for arbitrary values in $0\leq \Omega_{\Lambda, \text{M}}\leq 1$ is more complicated since it depends on several conditions on the roots of a cubic equation which arises while performing the integration and it is not guaranteed to be the same as that in \eqref{int-MF}.} and is given by
\begin{equation}\label{int-MF}
\mathcal M_\text{F}(z) \equiv \int_{r_i}^r dr^\prime M_\text{F}(r^\prime)=\frac{2}{3 \Omega_\text{M}}\mathcal A_1 \mathcal A_2 H_0^{-1} \sin(\Theta)\sin(\Phi)\left(\sqrt{\Omega_\Lambda +\Omega_\text{M}(1+z_i)^3} - \sqrt{\Omega_\Lambda +\Omega_\text{M}(1+z)^3}\right),
\end{equation}
where $\mathcal A_2 \equiv 2.8\times 10^{6} (\bar B_0/\text{G})(\text{Hz}/\nu_0)$. In the case of $\mathcal M_\text{C}$ exact expressions for any $z$ do not exist but only in some limiting cases
\begin{align}
\mathcal M_\text{C}(z) & \equiv \int_{r_i}^r dr^\prime M_\text{C}(r^\prime) \simeq \begin{cases} \sin(2\Theta) \cos(\Phi) (\mathcal A_1/5)\,\mathcal A_2^2\, \Omega_\text{M}^{-1/2}\,H_0^{-1} \left[\sqrt{1+z_i}\left(1+z_i(2+z_i) \right)-\sqrt{1+z}\left( 1+ z(2+z)\right)\right]\\ \text{for}\quad z\gg \left[(\Omega_\Lambda/\Omega_\textrm{M})^{1/3}-1\right], \\
 \sin(2\Theta) \cos(\Phi) (\mathcal A_1/8)\,\mathcal A_2^2\,\Omega_\Lambda^{-1/2}\,H_0^{-1}\,\left[ (z_i-z)\left(2+z_i+z)(2+z_i(2+z_i) + z(2+z)\right)\right]\\ \text{for}\quad z\ll \left[(\Omega_\Lambda/\Omega_\textrm{M})^{1/3}-1\right].
\end{cases}
\end{align}

\subsection{The case when $\Phi = \pi/2$.}
\label{subsec:6}

In this section we study the particular case when $\Phi = \pi/2$, which essentially corresponds to $\mathcal M_\text{C}(z) = 0$. In this case in $\mathcal M_\text{CF}$, only the Faraday effect term $\mathcal M_\text{F}(z)$ is present, which we assume to be different from zero, namely when $\sin(\Theta)\neq 0$. Indeed, if $\sin(\Theta)\neq 0$, the Faraday effect term is several orders of magnitude bigger than the CM effect without necessarily having the condition $\Phi = \pi/2$. Therefore the latter condition is a formal one as far as the Faraday effect term is different from zero. For $\Phi=\pi/2$, we have that $\mathcal C(r)= i$ and $\mathcal M_\text{CM}(r)= \mathcal M_\text{F}(r)$.

With the above considerations, let us now concentrate on the calculation of the integral $I_1$ in \eqref{integrals}, which for $\Phi = \pi/2$ becomes
\begin{align}\label{I1-1}
I_1(z) & = \int_{z}^{z_i}\,\frac{dz^{\prime}}{H_0(1+z^\prime)\sqrt{\Omega_\Lambda+\Omega_\text{M}(1+z^\prime)^3}} \cos\left[\mathcal M_\text{F}(z^\prime)\right] M_{g\gamma}^x(z^\prime) e^{- i  M_1(z^\prime)}\nonumber\\
& = \int_{z}^{z_i}\,\frac{dz^{\prime}}{H_0(1+z^\prime)\sqrt{\Omega_\Lambda+\Omega_\text{M}(1+z^\prime)^3}} \left( \cos\left[\mathcal M_\text{F}(z^\prime)\right] \cos[M_1(z^\prime)]M_{g\gamma}^x(z^\prime) - i\,\cos\left[\mathcal M_\text{F}(z^\prime)\right] \sin[M_1(z^\prime)]  M_{g\gamma}^x(z^\prime) \right)
\end{align}
Even though the integral $I_1$ has been significantly simplified for $\Phi = \pi/2$, it is still not possible to find an analytic expression because of the complexity of the integrands. Let us in addition assume that $\Theta\rightarrow 0$, which means that $\mathcal M_\text{F}\ll 1$; namely the external magnetic field is almost transverse with respect to the GW/electromagnetic wave propagation. In this regime, we can approximate $\cos[\mathcal M_\text{F}(z)]\simeq 1$ in \eqref{I1-1} and get
\begin{align}\label{I1-1-0}
I_1(z) & \simeq \int_{z}^{z_i}\,\frac{dz^{\prime}}{H_0(1+z^\prime)\sqrt{\Omega_\Lambda+\Omega_\text{M}(1+z^\prime)^3}} \left( \cos[M_1(z^\prime)]M_{g\gamma}^x(z^\prime) - i\, \sin[M_1(z^\prime)]  M_{g\gamma}^x(z^\prime) \right)
\end{align}
The integrals of the first and second terms in \eqref{I1-1-0} can be calculated exactly and are given by 
\begin{align}\label{I1-2}
\frac{\kappa}{2}\int_{z}^{z_i}\,\frac{dz^{\prime}}{H_0(1+z^\prime)\sqrt{\Omega_\Lambda+\Omega_\text{M}(1+z^\prime)^3}} \bar B(z^\prime) \cos[M_1(z^\prime)] &=  \mathcal C \,\sin[M_1(z)], \nonumber\\
 i \frac{\kappa }{2}\int_{z}^{z_i}\,\frac{dz^{\prime}}{H_0(1+z^\prime)\sqrt{\Omega_\Lambda+\Omega_\text{M}(1+z^\prime)^3}} \bar B(z^\prime) \sin[M_1(z^\prime)] &=  i\,\mathcal C \,\left(1- \cos[M_1(z)]\right).
\end{align}
where $\mathcal C \equiv 9.75\times 10^{-3}\,\mathcal A_1^{-1}\,\kappa\,(\bar B_0/\text{G})$ (eV$^2$). Now we can use the expressions in \eqref{I1-2} in the integral in \eqref{I1-1-0} and obtain the final expression
\begin{equation}\label{I1-3}
I_1(z)= -\mathcal  C\,\left( i- i\cos[M_1(z)]- \sin[M_1(z)] \right).
\end{equation}
We may also note that in the limits where we found $I_1(z)$, we have that $I_1(z)=I_4(z)$. Again in this limit we have from \eqref{integrals} that $I_2(z)\simeq I_3(z)$. In addition, in the limits considered in this section, we have that $|I_1(z)|\gg |I_2(z)|$ since the integrand in $I_2$ is proportional to $\sin[\mathcal M_\text{F}] M_{g\gamma}^x\simeq \mathcal M_\text{F} M_{g\gamma}^x \ll 1$.

\section{Electromagnetic radiation from astrophysical binary systems}
\label{sec:7}

In the previous section we have been able to find analytic expressions for the integrals in \eqref{integrals} in the case when $\mathcal M_F\ll 1$ and $\Phi=\pi/2$. In more general cases it is not possible to find analytic solutions due to the complexity of the integrands in \eqref{integrals} ,and in these cases numerical results may be in order. In this section, we want to calculate some quantities related to the electromagnetic radiation in the GRAPH mixing such as the energy power $P_{\gamma}$ and/or the energy power flux $F_\gamma$. The latter quantity is simply given by $I_\gamma$ in \eqref{ph-int-1} where by definition $I_\gamma$ represent the energy density of photons at a given point in space, while the former quantity can easily be calculated once we know the distance of the GW source. 

The intensity of the generated electromagnetic radiation in the GRAPH mixing, in the case when $\mathcal M_F\ll 1$ and $\Phi=\pi/2$ and by using the results of the previous section, is given by
\begin{equation}\label{I1-4}
I_\gamma(r, t) \simeq \omega^2(t) |I_1(r)|^2 \left[  | h_\times(r_i, t_i)|^2 + | h_+(r_i, t_i)|^2 \right],
\end{equation}
where we have neglected the term proportional to $\text{Re}\{...\}$ in $I_\gamma$ in \eqref{ph-int-1} because it is a small quantity with respect to the other terms and we used the fact that $I_1=I_4$ in the limit $\Theta\rightarrow 0$ and $\Phi=\pi/2$. Therefore, in order to find the intensity of electromagnetic radiation or related quantities at given distance $r$, we need the amplitudes of the GW at the distance $r_i$ when GWs enter the region of magnetic field.

The amplitudes of GWs of binary systems of astrophysical sources are usually calculated starting from the multipole expansion of the stress energy-momentum tensor of the source. For binary systems, typically the quadrupole approximation of a quasicircular orbit is a rather good approximation up to a maximum frequency $\nu_\text{max}$ (see discussion section below), where beyond this frequency the strong gravity effects become dominant and the binary system coalesces. Therefore, let us assume that we have a binary system which emits GWs and which is undergoing an inspiral phase of quasicircular motion.  The amplitudes of GWs at a distance $r$ from the source in the quadrupole approximation and in the local wave zone are given by \cite{Maggiore:1900zz}
\begin{align}\label{GW-amplitudes}
\kappa h_+(r, t_s) & = h_c(t_s^\text{ret}) \left(\frac{1+\cos^2(\iota)}{2}\right)\cos[\Psi(t_s^{\text{ret}})], \quad \kappa h_\times(r, t_s)=h_c(t_s^\text{ret}) \cos(\iota)\sin[\Psi(t_s^\text{ret})],\nonumber \\
h_c(t_s^\text{ret}) & \equiv \frac{4}{r} \left(G_\text{N} M_\text{CH} \right)^{5/3} \left[\pi \nu_s(t_s^\text{ret})\right]^{2/3}, \quad \Psi(t_s^\text{ret})\equiv \int^{t_s^\text{ret}} dt_s^{\prime}\,\omega_s(t_s^{\prime})
\end{align}
where $t_s$ is the time measured in the reference system of the GW source, $t_s^\text{ret}=t_s-r$ is the retarded time, $M_\text{CH}=(m_1m_2)^{3/5}(m_1+m_2)^{-1/5}$ is the chirp mass of the source with $m_{1,2}$ being the mass components of the binary system, and $\iota$ is the angle of the normal of the binary system orbit with respect to the direction of observation. 
We may note the factor $\kappa$ in \eqref{GW-amplitudes} that we have introduced in order to conform with the notation used in Ref. \cite{Maggiore:1900zz}, which uses the metric expansion $g_{\mu\nu}=\eta_{\mu\nu}+ h_{\mu\nu}$, while in our notations we use $g_{\mu\nu}=\eta_{\mu\nu}+ \kappa h_{\mu\nu}$.

The GWs amplitudes in \eqref{GW-amplitudes} are expressed in terms of the source variables that are measured in the source reference system. Moreover, they do not take into account the universe expansion yet and have been calculated in the local wave zone, namely at distances $r\gg d$ where $d$ is the typical size of the binary system orbit. To make our treatment as  simple as possible, let us assume that at the initial distance $r_i$, in the local wave zone, is present a large-scale magnetic field. Let $r_0$ be the light traveled distance from the source until present epoch. Thus the effective distance traveled by GWs once they enter the region of large-scale magnetic field is $r_0-r_i\simeq r_0$ where $r_0\gg r_i$. It is more convenient for our purposes to express the amplitudes in \eqref{GW-amplitudes} in terms of laboratory variables at the present epoch. Consequently, we can write $\nu_s(t_s^\text{ret})=\nu_0(t_\text{0}^\text{ret})(1+z)$ where $t_\text{0}^\text{ret}=(1+z)t_s^\text{ret}$ is the observed retarded time in the laboratory reference system. One can also easily check that $\Psi(t_s^\text{ret})=\Psi(t_\text{0}^\text{ret})$. Therefore, the initial GW amplitudes that enter the region of large-scale magnetic field at the initial distance $r_i$ from the source, expressed in terms of present epoch variables, are given by
\begin{align}\label{GW-amplitudes-1}
\kappa h_+(r_i, t_\text{0}) & = h_c(t_\text{0}^\text{ret}) \left(\frac{1+\cos^2(\iota)}{2}\right)\cos[\Psi(t_\text{0}^{\text{ret}})], \quad \kappa h_\times(r_i, t_\text{0})=h_c(t_\text{0}^\text{ret}) \cos(\iota)\sin[\Psi(t_\text{0}^\text{ret})],\nonumber \\
h_c(t_\text{0}^\text{ret}) & \equiv \frac{4}{r_i} \left(G_\text{N} M_\text{CH} \right)^{5/3} \left[\pi \nu_0(t_\text{0}^\text{ret})\right]^{2/3}(1+z_i)^{2/3}.
\end{align}

At this stage there are two important things to point out, which are of great importance in what follows. So far, we have considered the propagation of the GWs in a magnetized plasma, and the equation of motion that we have derived in \eqref{schr-eq} takes into account the change of the initial GW amplitude in the GRAPH mixing only. However, for point sources of GWs, the amplitudes have an intrinsic decay with the distance of the form $\propto 1/r$. Intentionally we did not look for solutions of the form $h_{\times, +}(r, t)\propto 1/r$ and $A_{x, y}(r, t)\propto 1/r$ in Eqs. \eqref{GW-eq} and \eqref{PH-eq} in order to simplify our formalism as much as possible. So,  to include the intrinsic decay of the amplitudes with the distance in the expression of the intensity given in \eqref{I1-4}, we introduce the scaling $I_\gamma\rightarrow I_\gamma (r_i/r)^2$. Another important thing to note is that Eqs. \eqref{GW-eq} and \eqref{PH-eq} have been derived in  Minksowski spacetime. However, our problem of GRAPH mixing essentially needs to be applied to the FRW metric in the case when GWs propagate in an expanding universe. As shown in Ref. \cite{Ejlli:2016avx}, the universe expansion is represented by the Hubble friction term $-3 H\partial_t$ and if one includes this term in the equations of motion, the amplitude square of GWs ($h_{\times, +}$) and of electromagnetic radiation $(A_{x, y})$, scale with the redshift as $\propto (1+z)^{2}$. Since $I_\gamma\propto \omega^2 |A|^2$ represents the energy density of photons and because $\omega^2(z) \propto (1+z)^2$, we have that $I_\gamma(r_0, t_0) \propto (1+z)^4$. Consequently, we have that the intensity of the electromagnetic radiation at present, $t=t_0$ or $z=0$, is given by 
\begin{equation}\label{I1-5}
I_\gamma(r_0, t_0) \simeq \omega_0^2 |I_1(0)|^2 \left[  | h_\times(r_i, t_0)|^2 + | h_+(r_i, t_0)|^2 \right](1+z_i)^4(r_i/r_0)^2,
\end{equation}
where we remind the reader that $z_i$ is the redshift of the GW source at the present epoch which is not related to $r_i$. Here we are assuming that the redshift of the source $z_i$ is approximately the same as the redshift when GWs enter the region of the large-scale magnetic field.

The expression for the intensity in \eqref{I1-5} still is not in the final form because of the presence of $\sin[\Psi]$ and $\cos[\Psi]$ in the initial GW amplitudes and also because of the dependence on the angle $\iota$. At this point it is more convenient to average the intensity $I_\gamma$ over the phase $0\leq \Psi\leq 2\pi$ and $0\leq \iota\leq \pi$.  By putting all together, we get
\begin{align}\label{Av-int}
\bar I_\gamma(r_0, t_0) & \simeq \frac{35}{64}\left(\frac{2\pi\nu_0}{\kappa r_0}\right)^2 [r_i h_c(t_0^\text{obs})]^2\, |I_1(0)|^2 (1+z_i)^4\nonumber\\
& = \frac{35}{16}\left(\frac{2\pi\nu_0}{\kappa r_0}\right)^2 [r_i h_c(t_0^\text{obs})]^2\,  \mathcal C^2 \sin^2[M_1(0)/2](1+z_i)^4
\end{align}
where $\bar I_\gamma$ is the average value of the intensity on $\Psi$ and $\iota$ and \emph{not} on $\Phi$ and $\Theta$. The energy per unit time (or the power $P_\gamma$) of the electromagnetic radiation, generated in the GRAPH mixing is given by
\begin{align}\label{Av-pow}
\bar P_\gamma(t_0) & =4\pi r_0^2 \bar I_\gamma(r_0, t_0)=\frac{35\pi}{4}\left(\frac{2\pi\nu_0}{\kappa }\right)^2 [r_i h_c(t_0^\text{obs})]^2\,  \mathcal C^2 \sin^2[M_1(0)/2](1+z_i)^4\nonumber\\
& \simeq 3.89\times 10^{13} \left(\frac{M_\text{CH}}{M_\odot}\right)^{10/3}\left(\frac{\nu_0}{\text{Hz}}\right)^{16/3}\left(\frac{\bar B_0}{\text{G}}\right)^2\sin^2[M_1(0)/2](1+z_i)^{16/3}\quad (\text{erg}/\text{s}),
\end{align}
where $M_\odot$ is the solar mass.

\begin{figure*}[h!]
\centering
\mbox{
\subfloat[\label{fig:Fig1}]{\includegraphics[scale=0.67]{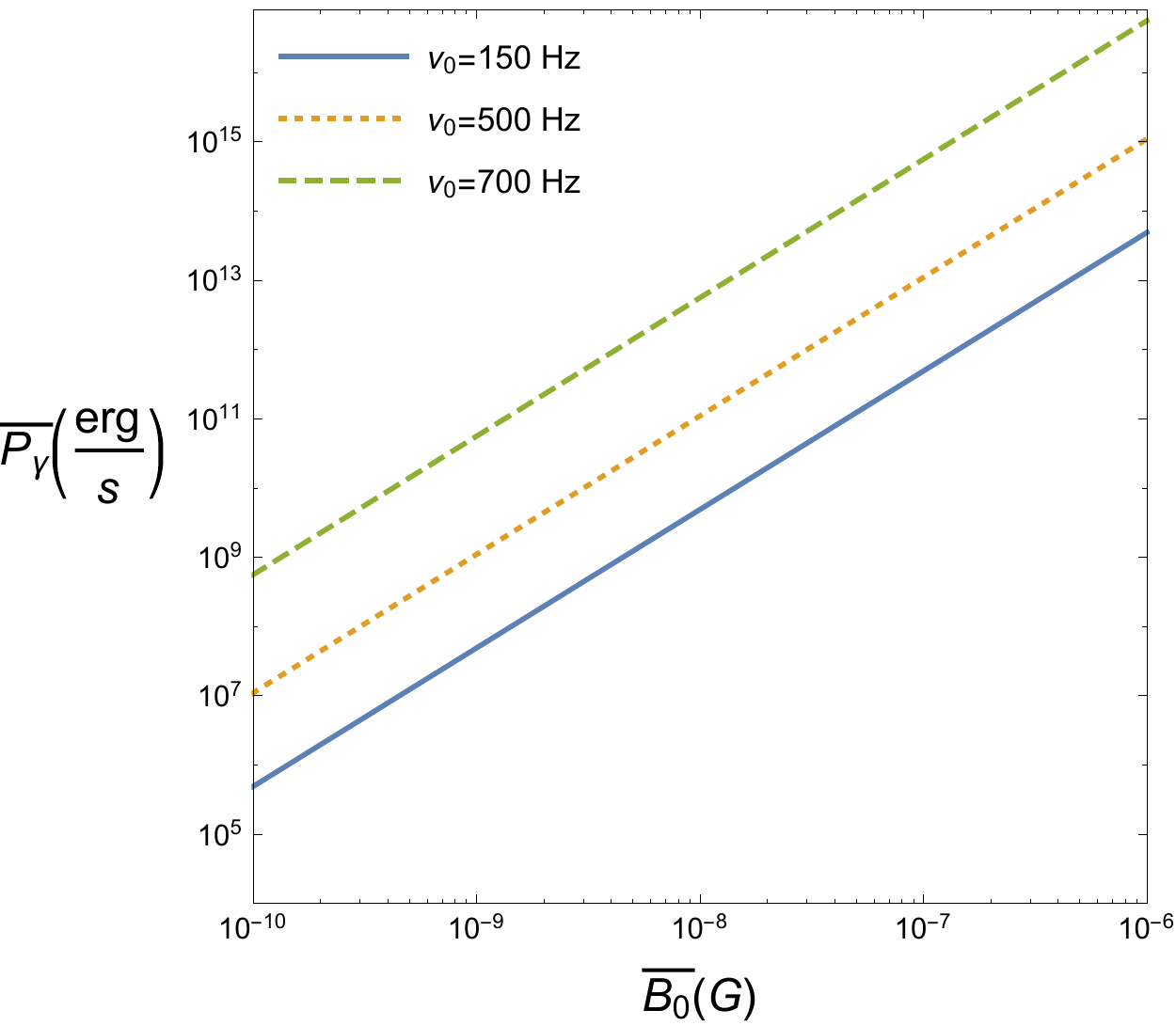}}\qquad
\subfloat[\label{fig:Fig2}]{\includegraphics[scale=0.66]{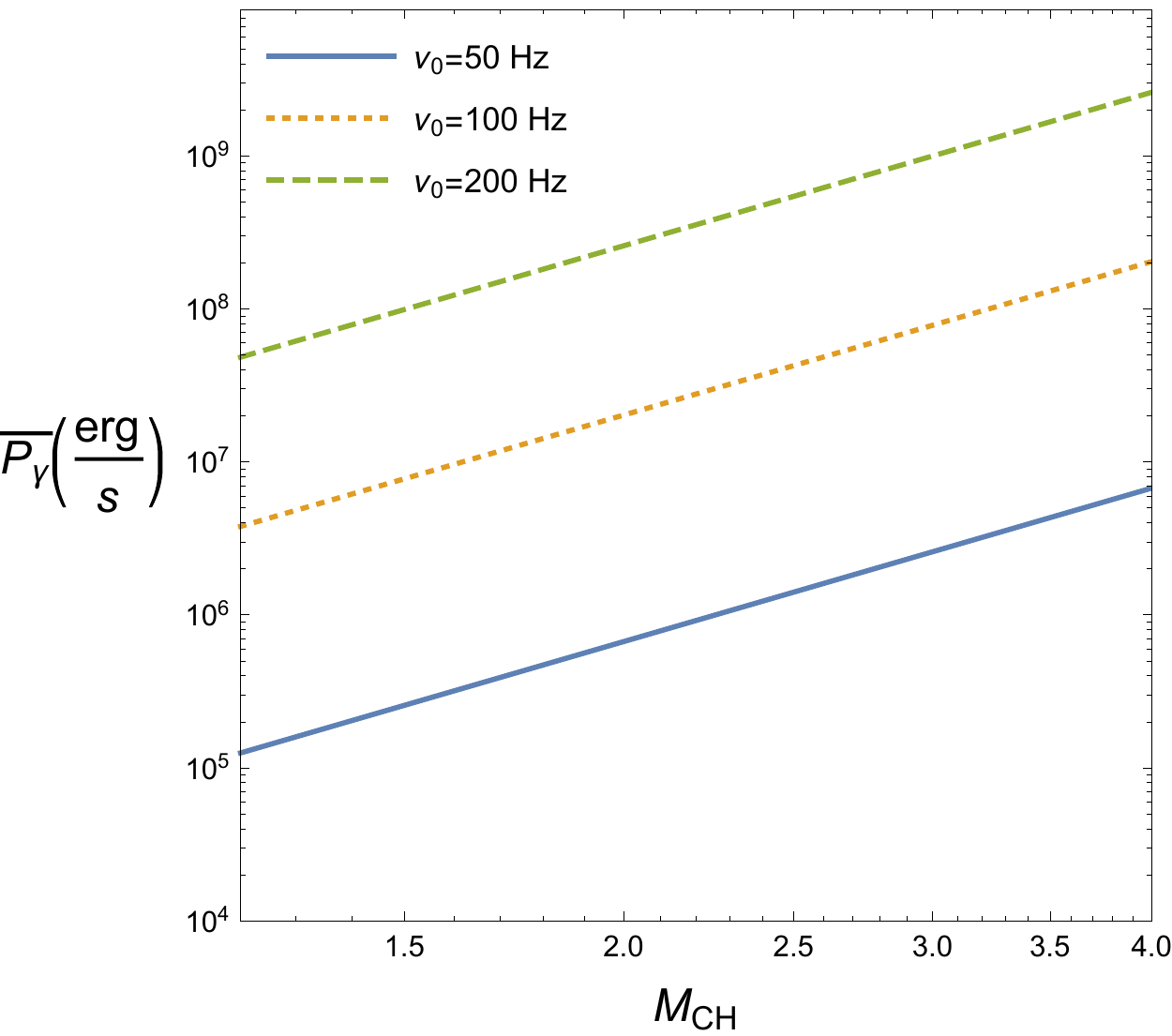}}}
\caption{(a) Logarithmic scale plots of the power of the electromagnetic radiation $\bar P_\gamma$ (erg/s) at the present time as a function of the present day value of cosmic magnetic field amplitude $\bar B_0 $ (G), generated in the GRAPH mixing, for a typical binary system of neutron stars with equal masses $m_1=m_2=1.4 M_\odot$ and chirp mass $M_\text{CH}\simeq 1.21 M_\odot$, for $z=0.1$ and frequencies $\nu_0=\{150, 500, 700\}\, \text{Hz} $ are shown. (b) Logarithmic scale plots of the power of the electromagnetic radiation $\bar P_\gamma$ (erg/s) at present time as a function of the binary system chirp mass $M_\text{CH}$ (in units of the solar mass) for a binary system of equal masses, for $\bar B_0=1$ nG, $z=0.1$, and frequencies $\nu_0=\{50, 100, 200\}\, \text{Hz} $ are shown.}
\label{fig:Fig1a}
\end{figure*}

\begin{figure*}[h!]
\centering
\mbox{
\subfloat[\label{fig:Fig3}]{\includegraphics[scale=0.65]{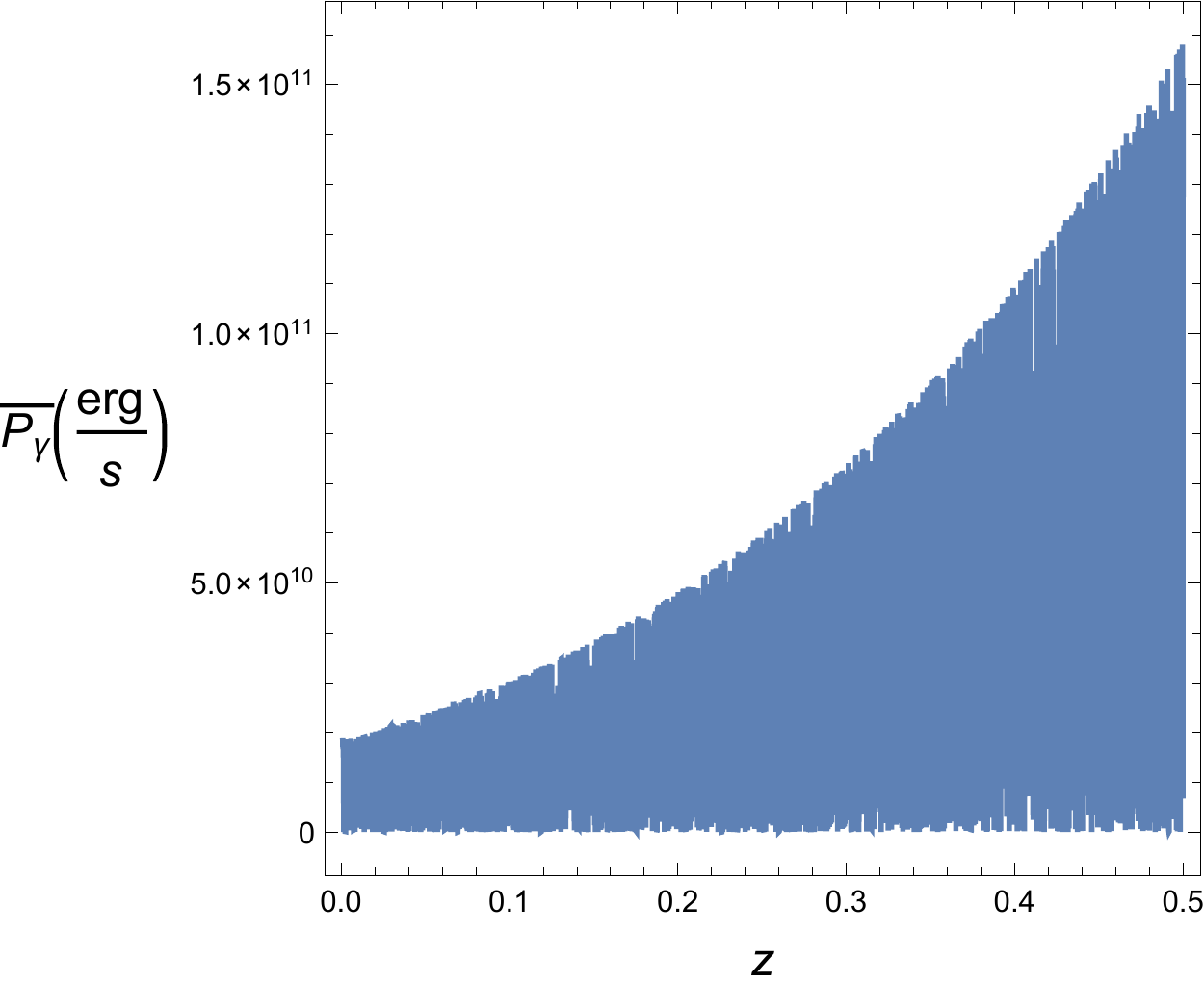}}\qquad
\subfloat[\label{fig:Fig4}]{\includegraphics[scale=0.635]{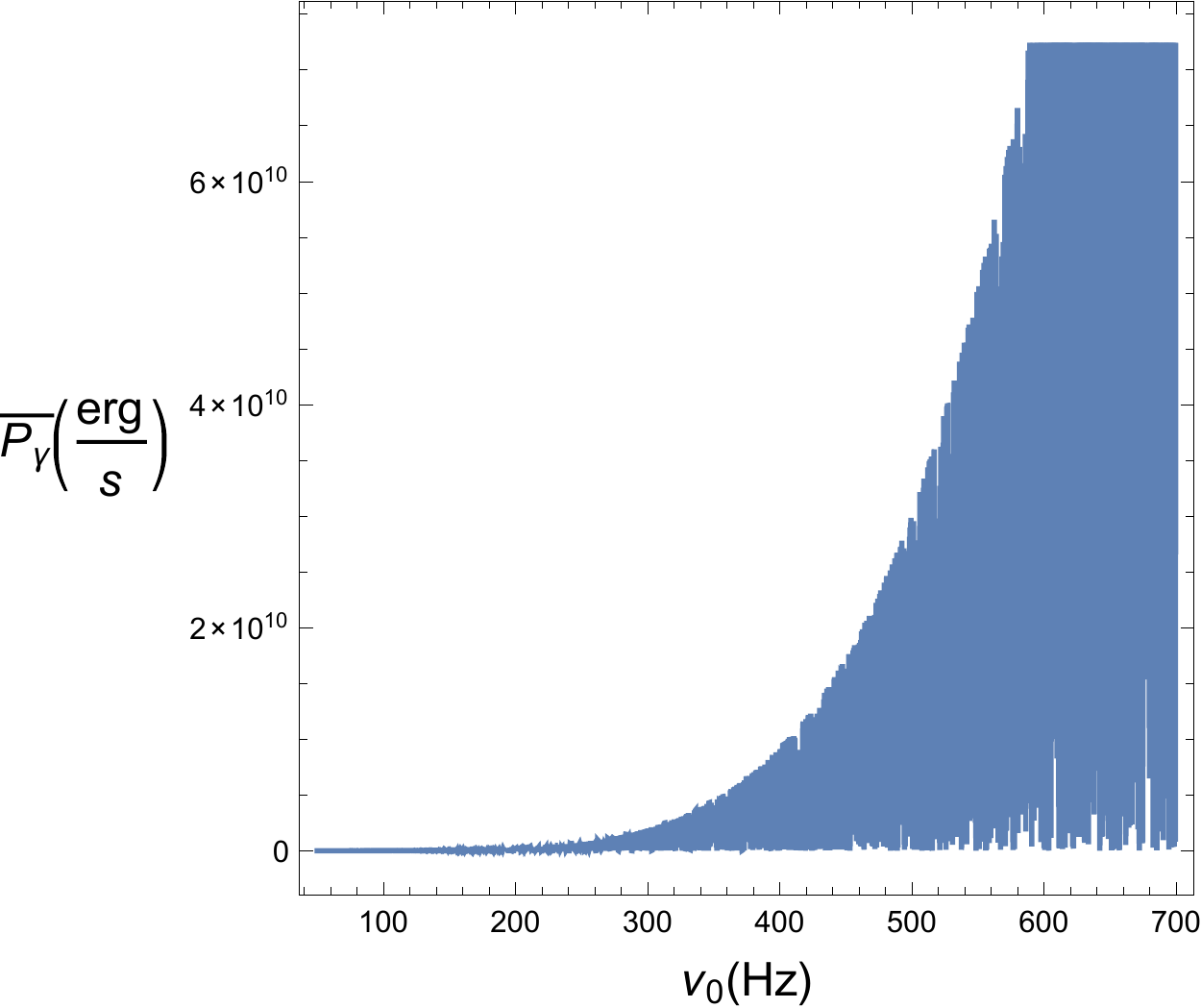}}}
\caption{(a) The power of the electromagnetic radiation $\bar P_\gamma$ (erg/s) at present time as a function of the GW source redshift $z\in[10^{-3}, 0.5]$, generated in the GRAPH mixing, for a typical binary system of neutron stars with equal masses $m_1=m_2=1.4 M_\odot$ and chirp mass $M_\text{CH}\simeq 1.21 M_\odot$, for $\bar B_0=1$ nG and frequency $\nu_0= 500\, \text{Hz} $ is shown. (b) The power of the electromagnetic radiation $\bar P_\gamma$ (erg/s) at present time as a function of the GW frequency $\nu_0\in[50, 700]$ Hz for a binary system with equal masses $m_1=m_2=1.4 M_\odot$ and chirp mass $M_\text{CH}=1.21 M_\odot$, for $\bar B_0=1$ nG and $z=0.1$ is shown.}
\label{fig:Fig2a}
\end{figure*}

In Fig. \ref{fig:Fig1a} plots of the average power of electromagnetic radiation, given in \eqref{Av-pow},  generated in the GRAPH mixing are shown. In Fig. \ref{fig:Fig1} the plots of the power as a function of the present day value of the cosmological magnetic field are shown. We may note that the power emitted is proportional to $\nu_0^{16/3}$ and also is  proportional to $\sin^2[M_1(0)/2]$. Thus, even though for higher values of the frequencies $\nu_0^{16/3}$ increases, it is also true that $\sin^2[M_1(0)/2]$ is an extremely oscillating function of the frequency, and consequently higher values of the frequency do not necessarily imply higher values of the power. The fast oscillatory behaviour of the average power as a function of the frequency and redshift, due to the term $\sin^2[M_1(0)/2]$, are shown in Fig. \ref{fig:Fig2a}.

\begin{figure*}[h!]
\centering
\mbox{
\subfloat[\label{fig:Fig5}]{\includegraphics[scale=0.65]{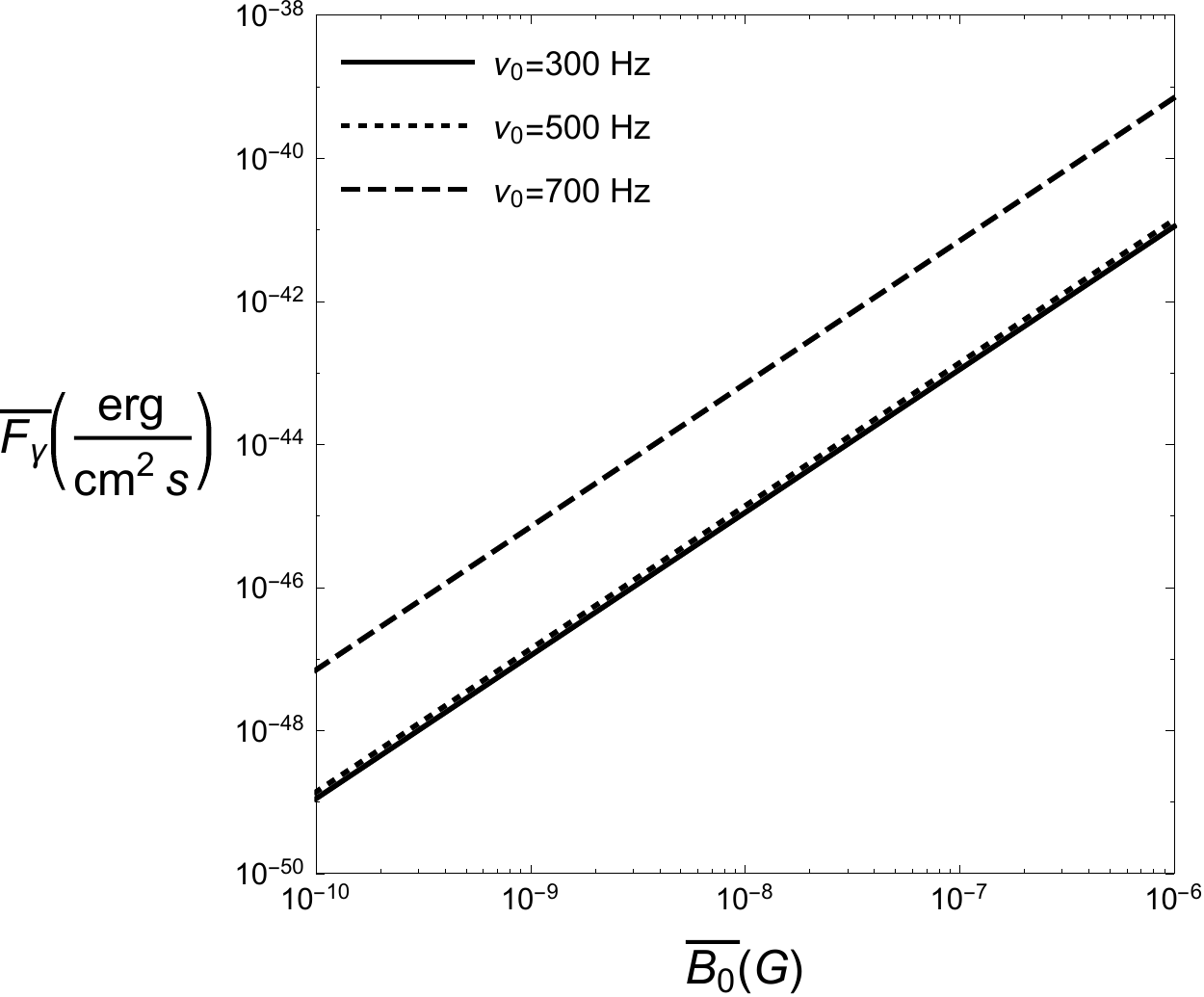}}\qquad
\subfloat[\label{fig:Fig6}]{\includegraphics[scale=0.67]{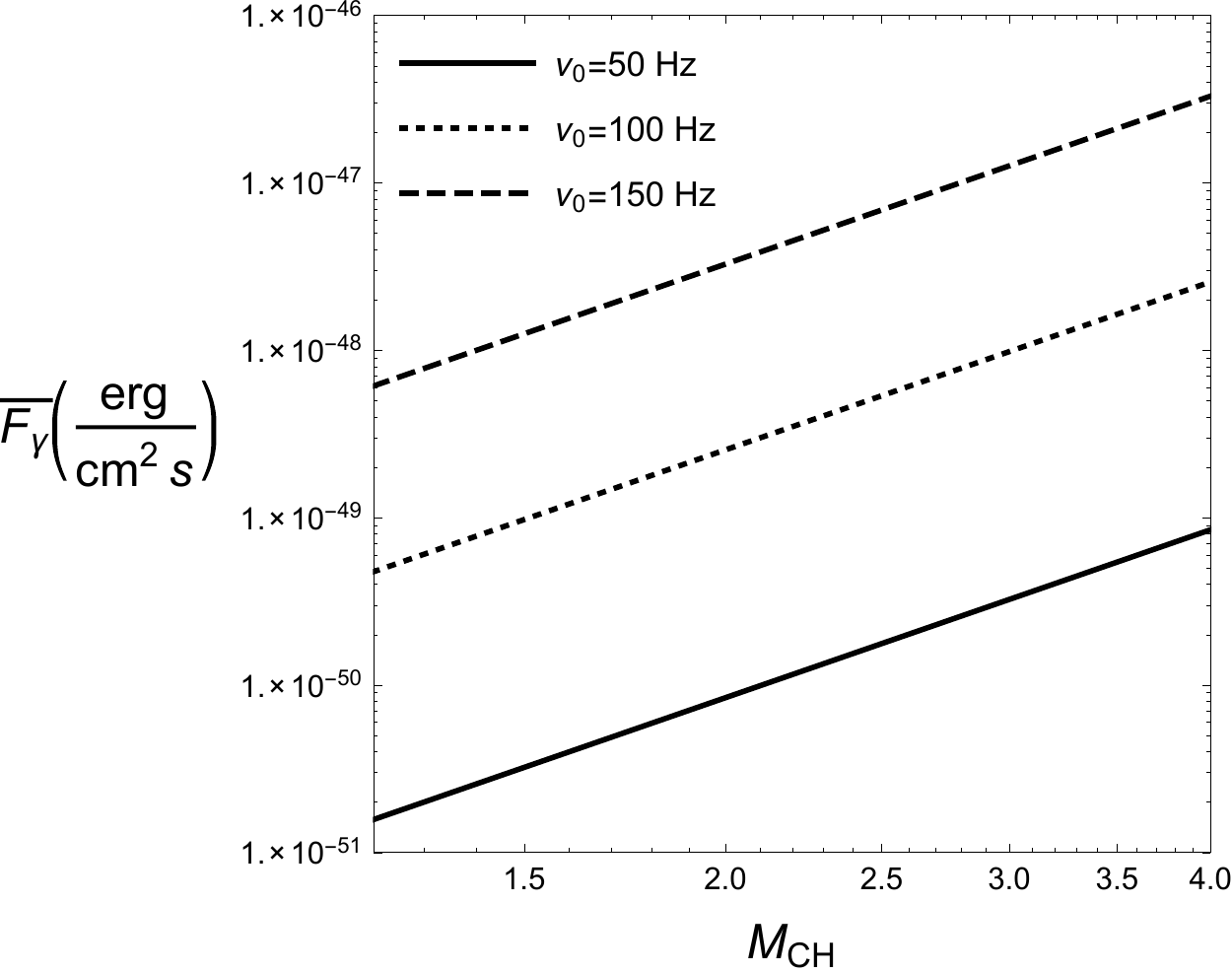}}}
\caption{a) Logarithmic scale plots of the power fluxes of the electromagnetic radiation $\bar F_\gamma$ (erg cm$^{-2}$ s$^{-1}$) at present time as a function of the present day value of the magnetic field $\bar B_0\in[10^{-10}, 10^{-6}]$ (G), generated in the GRAPH mixing, for a typical binary system of neutron stars with equal masses $m_1=m_2=1.4 M_\odot$ and chirp mass $M_\text{CH}\simeq 1.21 M_\odot$, for a source located at redshift $z=0.1$ and frequencies $\nu_0= \{300, 500, 700\}\,\text{Hz} $ are shown. (b) Logarithmic scale plots of the power fluxes of the electromagnetic radiation $\bar F_\gamma$ (erg cm$^{-2}$ s$^{-1}$) at present time as a function of the source chirp mass $M_\text{CH}\in [1.21, 4] M_\odot$ for $\bar B_0=1$ nG, source redshift $z=0.1$, and frequencies $\nu_0= \{50, 100, 150\}\,\text{Hz} $ are shown.}
\label{fig:Fig3a}
\end{figure*}

\begin{figure*}[h!]
\centering
\mbox{
\subfloat[\label{fig:Fig7}]{\includegraphics[scale=0.65]{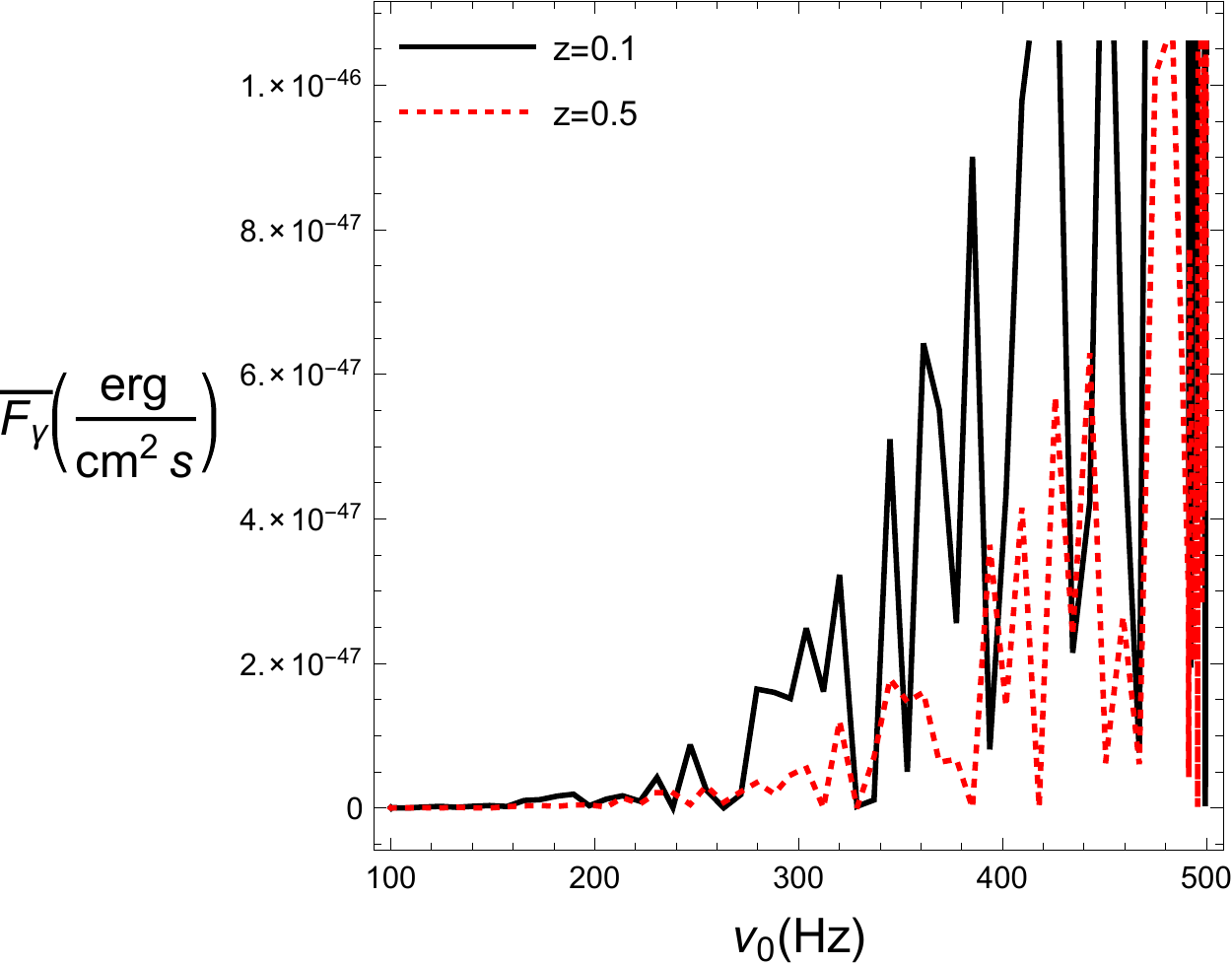}}\qquad
\subfloat[\label{fig:Fig8}]{\includegraphics[scale=0.67]{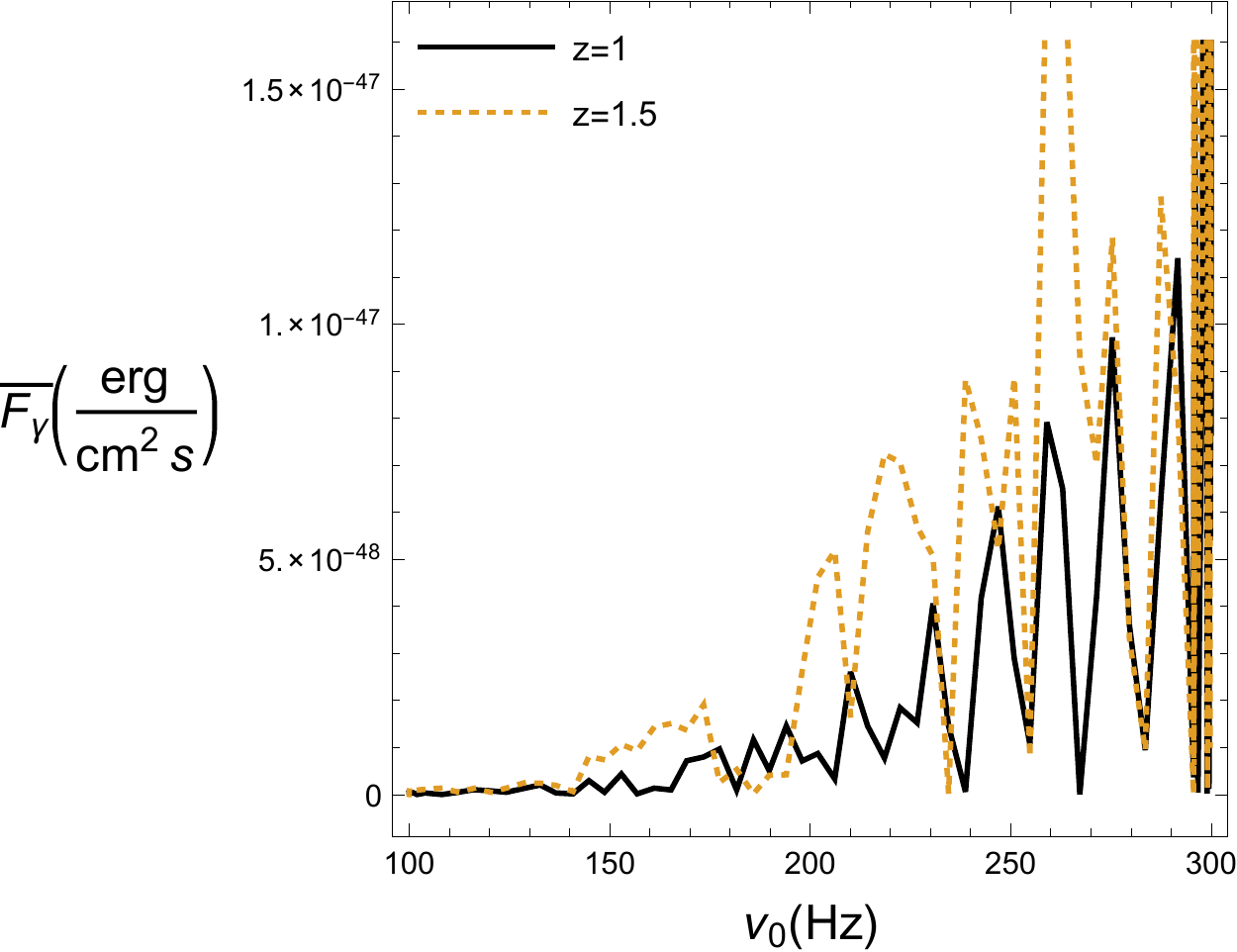}}}
\caption{(a) The power fluxes of the electromagnetic radiation $\bar F_\gamma$ (erg cm$^{-2}$ s$^{-1}$) at present time as a function of the present day value of GW frequency $\nu_0 \in[100, 500]$ (Hz), generated in the GRAPH mixing, for a typical binary system of neutron stars with equal masses $m_1=m_2=1.4 M_\odot$ and chirp mass $M_\text{CH}\simeq 1.21 M_\odot$, for $\bar B_0=1$ (nG) and source redshifts $z= 0.1$ and $z= 0.5$ are shown. (b) Similar plots as in (a) for source redshifts $z=1$ and $z=1.5$ and frequency interval $\nu_0 \in[100, 300]$ (Hz) are shown.}
\label{fig:Fig4a}
\end{figure*}

In Figs. \ref{fig:Fig3a}-\ref{fig:Fig4a} the average power fluxes of the electromagnetic radiation generated in the GRAPH mixing received at Earth, given by expression \eqref{Av-int}, for a source of GWs located at redshift $z_i$ are shown. As we can see, the average power fluxes received today are quite faint, and they rapidly oscillate with the frequency $\nu_0$ and the redshift $z$. The rapid oscillation of the received energy power flux is evident in Fig. \ref{fig:Fig4a}, where plots of the energy power flux as a function of the frequency are shown. As already discussed above in the case of the energy power, higher values of the frequencies do not necessarily mean higher values of the power flux. Again, this behavior is due to the $\sin^2[M_1(0)/2]$ term in \eqref{Av-int}, which is an extremely fast oscillating function of the frequency. In addition, as we can see in Fig. \ref{fig:Fig8}, there are cases where the average energy power flux received from closer to the Earth binary systems or low redshift, $z$, GW sources, is smaller than the average energy power flux received from far away binary systems of higher redshifts. This behaviour is still due to the factor  $\sin^2[M_1(0)/2]$, which explicitly depends on the redshift and consequently is an extremely fast oscillating function of $z$ as well.

As already discussed above and shown in Figs. \ref{fig:Fig1a}-\ref{fig:Fig4a}, the energy power and energy power flux given, respectively, in expressions \eqref{Av-pow} and \eqref{Av-int} are proportional to the term $\sin^2[M_1(0)/2]$, which is an extremely fast oscillating function of the parameters. It may be convenient for several reasons to average the energy power and energy power fluxes over a given observation frequency range. This might be, for example, the case of a detector that measures the energy power flux in a specific frequency range due to the detector characteristics.  In this case, we have to average $\nu_0^{16/3}\sin^2[M_1(0)/2]$ on a frequency interval. However, since the integral of $\nu_0^{16/3}\sin^2[M_1(0)/2]$ is not an elementary one, for simplicity here we make the following observation; given a frequency interval with $0<\nu_{0, 1}\leq \nu_0\leq {\nu_{0, 2}}$ (where the frequency is expressed in units of Hz), we have that
\begin{equation}\nonumber
\left| \int_{\nu_{0, 1}}^{\nu_{0, 2}} d\nu^\prime \nu_{0}^{\prime 16/3}\sin^2[M_1(0)/2] \right|\leq \int_{\nu_{0, 1}}^{\nu_{0, 2}} d\nu^\prime \left| \nu_{0}^{\prime 16/3}\sin^2[M_1(0)/2] \right| \leq \int_{\nu_{0, 1}}^{\nu_{0, 2}} d\nu^\prime \left| \nu_{0}^{\prime 16/3} \right|=\frac{3}{19}\left( \nu_{0, 2}^{19/3} - \nu_{0, 1}^{19/3}\right).
\end{equation}
Consequently, we have, for example that the average value on the frequency of the energy power is at maximum

\begin{align}\label{Av-pow-1}
\langle \bar P_\gamma(t_0) \rangle & \simeq 3.89\times 10^{13} \left(\frac{M_\text{CH}}{M_\odot}\right)^{10/3}\left(\frac{\bar B_0}{\text{G}}\right)^2 \left\langle \nu_0^{16/3}\sin^2[M_1(0)/2] \right\rangle (1+z_i)^{16/3} \quad (\text{erg}/\text{s}) \nonumber\\
& \leq 1.3\times 10^{13} \left(\frac{M_\text{CH}}{M_\odot}\right)^{10/3}\left(\frac{\bar B_0}{\text{G}}\right)^2  (1+z_i)^{16/3}\left( \frac{ \nu_{0, 2}^{19/3} - \nu_{0, 1}^{19/3}}{ \nu_{0, 2} - \nu_{0, 1}}\right)      \quad (\text{erg}/\text{s}).
\end{align}
For example, if we consider a GW source with $M_\text{CH}=1.21 M_\odot, \bar B_0=1 \text{nG}$ and $\nu_{0, 1}=50$ Hz, $\nu_{0, 2}=600$ Hz, we get from \eqref{Av-pow-1}
\begin{align}\label{Av-pow-2}
\langle \bar P_\gamma(t_0) \rangle & \leq 1.75\times 10^{10}  (1+z_i)^{16/3}   \quad (\text{erg}/\text{s}) \quad \text{for} \quad z\ll 1.
\end{align}
It is very important to stress that we are considering values of redshifts in order that the GWs frequencies, which we consider in our calculations and plots, must be below or at maximum equal to the ISCO frequency, $(\nu_0)_\text{ISCO}$, for given values of $M_\text{T}$; see Sec. \ref{sec:9}.

At this point we can calculate from \eqref{Av-pow-1} and \eqref{Av-pow-2} the upper limit of the frequency averaged power flux which is given by $\langle F_\gamma(r_0, t_0)\rangle =\langle \bar P_\gamma(t_0) \rangle/(4\pi r_0^2)$. We can calculate the light traveled distance $r_0$ from the expression \eqref{light-t-distance}. However, since does an analytic expression for the integral in \eqref{light-t-distance} does not exist for arbitrary $z$, let us consider for simplicity the case of low redshift GW sources, such as $z\ll \left[(\Omega_\Lambda/\Omega_\textrm{M})^{1/3}-1\right]\simeq 0.29$. In this case we find $r_0-r_i\simeq r_0 \simeq 3.25\times 10^{28}\ln(1+z_i)$ (cm). Therefore, the upper limit of the frequency averaged value of the power flux at $r_0$ and $t_0$ is given by
\begin{equation}\label{Av-int-1}
\langle \bar F_\gamma(r_0, t_0) \rangle \leq 9.8\times 10^{-46} \left(\frac{M_\text{CH}}{M_\odot}\right)^{10/3}\left(\frac{\bar B_0}{\text{G}}\right)^2 \left( \frac{(1+z_i)^{16/3}}{\ln^2(1+z_i)} \right) \left( \frac{ \nu_{0, 2}^{19/3} - \nu_{0, 1}^{19/3}}{ \nu_{0, 2} - \nu_{0, 1}}\right)      \quad (\text{erg}\; \text{s}^{-1} \text{cm}^{-2}).
\end{equation}
If we take again, for example, the same parameters as above, namely $M_\text{CH}=1.21 M_\odot, \bar B_0=1 \text{nG}$ and $\nu_{0, 1}=50$ Hz, $\nu_{0, 2}=600$ Hz, we get from \eqref{Av-int-1}
\begin{equation}\label{Av-int-2}
\langle \bar F_\gamma(r_0, t_0) \rangle \leq 1.32\times 10^{-48} \left( \frac{(1+z_i)^{16/3}}{\ln^2(1+z_i)} \right)\quad  (\text{erg}\; \text{s}^{-1} \text{cm}^{-2}) \quad \text{for} \quad z\ll 0.29.
\end{equation}


\section{Frequency cut-offs and detectability}
\label{sec:8}

So far, in our analysis we considered the propagation of GWs in intergalactic magnetic fields for extragalactic binary systems of GWs, namely for a given binary system outside our galaxy. As shown in our plots, we considered binary systems with redshifts of $0.1 \leq z$ that could be located in any direction with respect to the observer and that might have formed after formation of first starts and also applicable to black hole binary systems with primordial origin. Therefore, the distances of these objects are usually a fraction of the present day Hubble distance $H_0^{-1}$ with distances equal to or larger than a few Gpc. The propagation of GWs from the source to the detector usually can be divided into three parts: GWs propagate from the source into the galactic medium that hosts the source, then they propagate into the intergalactic medium, and at the end they propagate inside our galaxy, the Milky Way.

In general the propagation of GWs from the source to the detector through galaxies and intergalactic space is rather complicated to model. Within the host galaxy where the source is located and in the local zone approximation $r\gg d$, where $r$ could even be outside the host galaxy depending where the source is located within the host galaxy, the generation of electromagnetic radiation is negligible (see below) with respect to the case when GWs enter the intergalactic space region. This happens because the number density of free electrons in galaxies is much larger (also the plasma frequency) than in intergalactic space and because the propagation distance in galaxies is smaller than in intergalactic space. When GWs propagate through the intergalactic space where a large-scale cosmic magnetic field might exist, the GRAPH mixing starts taking place efficiently since the number density of free electrons in void regions is expected to be $n_e(T) \simeq 0.76\, n_B(T) X_e(T)$, namely a fraction of the total baryon density as discussed in Sec. \ref{sec:3}, and the distance traveled is bigger than in galaxies.

When the formed electromagnetic radiation in the GRAPH mixing enters our galaxy, the number density of free electrons is much larger than in the intergalactic space, and it has been observed to be in the range $10^{-4}$ cm$^{-3}$ $\leq n_e(T_0)\leq 0.1$ cm$^{-3}$ in the interstellar medium depending on the line of sight. In the interstellar medium of our galaxy and in the case when $n_e\simeq 10^{-4}$ cm$^{-3}$, the plasma frequency along the line of sight would be (see Sec. \ref{sec:3}) $\nu_\text{pl}=8976.33\sqrt{n_e/\text{cm}^3}$ Hz=89.76 Hz. On the other hand, if $n_e=0.1$ cm$^{-3}$ along the light of sight, then the corresponding plasma frequency would be $\nu_\text{pl} \simeq 2.84$ kHz. 
These calculations suggest that in the case when the formed electromagnetic radiation in the GRAPH mixing enters the interstellar medium in our galaxy in a region where $n_e\simeq 10^{-4}$ cm$^{-3}$ along the line of sight, only the electromagnetic radiation with $\nu_0> \nu_\text{pl}=89.76$ Hz will propagate and the electromagnetic radiation with $\nu_0\leq 89.76$ Hz will be absorbed by the plasma. In the other extremum, if $n_e\simeq 0.1$ cm$^{-3}$ along the line of sight, only the electromagnetic radiation with $\nu_0> \nu_\text{pl}=2.84$ kHz will propagate. 

The electromagnetic radiation generated in the GRAPH mixing once has traveled through the interstellar space, escapes plasma absorption, and reaches our Solar system that has a variable free electron density as well. In the interplanetary space the plasma frequency is about $\nu_\text{pl}\simeq 20-40$ kHz while on Earth ionosphere is about $\nu_\text{pl}\simeq$ 10 MHz. In the context of this work, based on the considerations above, one should treat with care the results obtained because some part of the electromagnetic radiation generated in the GRAPH mixing for sources at cosmological distances might not propagate when it enters our galaxy, depending on the galactic free electron number density along the line of sight. The best way to detect this electromagnetic radiation would be in the space close to the boundary of the Solar system where the number density in the interstellar medium is lower with respect to interplanetary and Earth electron number densities. In this case, when $n_e \simeq 10^{-4}-10^{-3}$ cm$^{-3}$, only the corresponding electromagnetic radiation above 89 Hz and 300 Hz would propagate. If $n_e\simeq 0.01-0.1$ cm$^{-3}$ along the line of sight, only the corresponding electromagnetic radiation above 897 Hz and 2.84 kHz would propagate.

\section{Conclusions}
\label{sec:9}

In this work, we have studied the GRAPH mixing effect in a large-scale cosmic magnetic field and have applied it to the case of astrophysical GW sources. This effect, which has never been observed so far might have an important contribution to the electromagnetic radiation received from a binary system of a GW source. In this work, we considered all standard effects that generate dispersion and coherence breaking of the electromagnetic radiation generated in the GRAPH mixing. To obtain the energy power and energy power fluxes, we had to solve a system of linear differential equations with variable coefficients. To solve the equations of motion, we used the perturbation theory where 
the terms related to the interaction of GWs with electromagnetic waves in the mixing matrix $M$ have been considered as small perturbations with respect to dispersive and coherence breaking terms of the electromagnetic radiation.

From the technical point of view, even by using a perturbative approach to solve the equations of motion, the resulting final expressions for the Stokes parameters contain integrations on the redshift of complicated functions, and in most cases it is not possible to obtain analytic expressions of the integrals. Indeed, we have already seen this happen in Secs. \ref{sec:6} and \ref{sec:7} where we obtained analytic expressions for the integrals $I_{1, 2, 3, 4}$ only in the case where $\Phi=\pi/2$ and $\mathcal M_\text{F}\ll 1$. In more general cases where the angles $\Theta$ and $\Phi$ are different from zero, the integrals appearing in the Stokes parameters do not have analytic expressions, and to calculate the Stokes parameters, one must use numerical integration. In this work, we focused our attention on the $I_\gamma$ Stokes parameter and did not study the evolution of the polarization parameters $Q, U$, and $V$. However, it is quite evident from the expressions \eqref{ph-int-1} that the generated electromagnetic radiation in the GRAPH mixing is elliptically polarized; namely, it has both linear and circular polarizations.  

Our main goal in this work has been to obtain useful quantities such as the energy power and energy power fluxes of the electromagnetic radiation, which can be used in many contexts especially to confront with experimentally measurable quantities. In this regard, in Secs. \ref{sec:6} and \ref{sec:7}, we calculated the energy power $P_\gamma(t_0)$ and the energy power flux $F_\gamma(r_0, t_0)$ in the case of quasiperpendicular external magnetic field with respect to the GW direction of propagation, namely the case when $\Phi=\pi/2$ and $\mathcal M_\text{F}\ll 1$. In this regime, where analytic expressions do exist for $P_\gamma(t_0)$ and $F_\gamma(r_0, t_0)$, we have shown in Figs. \ref{fig:Fig1a} - \ref{fig:Fig4a} the power and power fluxes as a function of different quantities such as $\bar B_0, \nu_0, M_\text{CH}$ and the redshift $z$. The energy power $P_\gamma$ generated in the GRAPH mixing effect is usually quite substantial and in the interval $P_\gamma \simeq 10^6 - 10^{15}$ (erg/s) for magnetic field amplitudes $\bar B_0\in [10^{-10}, 10^{-6}]$ G. On the other hand, the energy power flux received on Earth is usually quite faint, and it depends on the distance of the source if other parameters are fixed. One common feature of $P_\gamma$ and $F_\gamma$ is that they are extremely fast oscillating functions of the frequency $\nu_0$ and redshift $z$. These features are, respectively, shown in Figs. \ref{fig:Fig2a} and \ref{fig:Fig4a}. The high oscillatory feature of $P_\gamma$ and $F_\gamma$, often makes it quite difficult to numerically average them over $\nu_0$. Indeed, since the function $\sin^2[M_1(0)/2]$ that does appear in \eqref{Av-int} and \eqref{Av-pow} is a highly oscillating one, it is necessary in many cases to keep several digits of accuracy in the argument in order to minimize calculation errors. Different levels of accuracy in the argument of $\sin^2[M_1(0)/2]$ may give slightly different values of $P_\gamma$ and $F_\gamma$ as functions of the parameters.

In the case when the direction of the cosmic magnetic field is arbitrary, it is not possible to find analytical expressions for $P_\gamma$ and $F_\gamma$ because of the complexity of the integrands in $I_{1, 2, 3, 4}$, which appear in the Stokes parameters. However, even though we did not calculate $P_\gamma$ and $F_\gamma$ in the general case, we can make some general discussions about their magnitudes. Indeed, by observing the integrands in the integrals in \eqref{integrals}, we may notice that in the general case when $\mathcal M_\text{CF}\neq 0$, the integrals $I_{1, 2, 3, 4}$ contain as integrands $\sin[\mathcal M_\text{CF}]$ and $\cos[\mathcal M_\text{CF}]$ multiplied with $M_{g\gamma}^{x, y} \sin[M_1]$ or $M_{g\gamma}^{x, y} \cos[M_1]$. Because of the fact that the absolute value of trigonometric functions is between zero and one, we expect that in the general case where $\mathcal M_\text{CF}\neq 0$, the magnitudes of $P_\gamma$ and $F_\gamma$, will be either smaller or at maximum the same as those found in the case when $\mathcal M_\text{CF}\rightarrow 0$ as explicitly calculated in Sec. \ref{sec:7}. Of course, this fact should not be a surprise since the Faraday and CM effects, appearing in $\mathcal M_\text{CF}$, are coherence braking and dispersive phenomena, which tend to limit the GRAPH mixing with respect to the case when these effects are almost absent.

There are several important points that deserve special discussions. First, in this work, we considered GWs with observed frequencies roughly speaking above 50 Hz and below 700 Hz. The reasons for this choice are strictly related to the approximations used in this work. In the lower frequency range, we considered GWs and electromagnetic waves with frequencies above the plasma frequency as discussed in Sec. \ref{sec:3}. If the GW frequency is below the plasma frequency, the electromagnetic wave generated in the GRAPH mixing would most likely not propagate in the plasma and be absorbed by it. This essential fact makes the GRAPH mixing less appealing for GWs with $\nu_0\lesssim $ few Hz. However, the common statement that the electromagnetic radiation does not propagate when the frequencies are below the plasma frequency is based on the assumption that external currents that couple to photons such as GWs in the macroscopic Maxwell equations, do not exist. But given the fact that such coupling is very small in general, we expect that the common statement that electromagnetic radiation with frequencies below that plasma frequency does not propagate, to remain still valid to the first order of approximation. On the other hand, we have chosen GWs emitted from binary systems in quasicircular motion in the quadrupole approximation. This approximation, as discussed in detail in Ref. \cite{Maggiore:1900zz}, is valid up to a maximum separation distance of the binary system that corresponds with maximum frequency equal to the present day ISCO frequency $\nu_0^\text{max}=(\nu_0)_\text{ISCO}\simeq 2.2\times 10^3(1+z_i)^{-1}(M_\odot/M_\text{T})$ Hz, where $M_\text{T}$ is the total mass of the binary system.

The second point is that the detection of the electromagnetic radiation in the GRAPH mixing on Earth and/or interplanetary space is very unlikely because of the large plasma frequency cutoffs as discussed above in Sec. \ref{sec:8}, and the best possibility would be to probe this signal beyond the Solar System. However, our analysis in this work has been done for binary systems undergoing quasicircular motion up to the maximum frequency $\nu_0^\text{max}=(\nu_0)_\text{ISCO}$ where the quasicircular motion approximation is valid. Obviously, the spectrum of GWs from these sources extends further even for $\nu_0\geq (\nu_0)_\text{ISCO}$ but the quasicircular motion approximation in this case is not valid. The GWs frequency spectrum for $(\nu_0)_\text{ISCO}\leq \nu_0\leq \nu_0^\text{coal}$ until the source coalescence is less likely to be absorbed in the interstellar medium because it is of high frequency and it may be possible to detect it even in the interplanetary space depending on the final coalescence frequency $\nu_0^\text{coal}$. The study of such possibility is beyond the scope of this work.

The third point is that we calculated the energy power and energy power flux only for a single binary system of GWs. While the energy power flux $P_\gamma$ is substantial, the energy power flux $F_\gamma$ is quite faint. However, since in the universe there are many sources of the stochastic background of GWs in every direction we expect that the energy power and energy power flux to be quite substantial in the interested frequency band. We plan to carry such a study in a forthcoming work and extend it to many sources of the stochastic background of GWs.

The fourth point is related to the fact that on deriving our results, we used only the three point GRAPH mixing, which is the lowest order of expansion in quantum field theory. One might wonder what are the consequences on the derived results in this work, if the one loop GRAPH mixing in the magnetic field is taken into account. As shown in Ref. \cite{Bastianelli05}, the one loop contribution to the GRAPH mixing in the magnetic field gives some correction to the usually studied three point GRAPH mixing transition amplitude, which depends nontrivially on the graviton/photon energy $\omega$ and magnetic field strength $B_e$. However, such a correction for weak magnetic fields, $B_e\ll B_c$, and for low energy gravitons/photons such as those considered in this work, is very small, and it can be safely neglected.

The fifth point is that in this work, we considered GWs generated in the quadrupole approximation which is valid for distances $r\gg d$ where $d$ is the typical size of the binary system. However, if $r< d$ and a magnetic field generated by internal process in the source in the binary system already exists, the GRAPH mixing effect can take place and generation of electromagnetic radiation might be substantial, given the fact that for binary systems of pulsars the magnetic field strength is very large and of the order of $B\simeq 10^{12}$ G. In any case, for $r<d$ the quadrupole approximation is not valid anymore and if a magnetic field exists at such distances, the effective GRAPH mixing strength is unknown because at such distances inside the GW source, usually there are five GW modes and not two as in the case of vacuum at distances $r\gg d$. The calculations of the GRAPH mixing strength for $r<d$ is beyond the purposes of this work.





\vspace{0.5cm}

 

  \end{document}